\documentclass[draft]{style_guide}
\usepackage{url} 
\usepackage{lineno}
\usepackage[inline]{trackchanges} 
\usepackage{soul}
\usepackage{amsmath}

\draftfalse

\journalname{ }

\begin{document}

\title{The Ensemble Kalman Inversion Race}


\authors{Rebecca Gjini\affil{1}, Matthias Morzfeld\affil{1}, Oliver R.A. Dunbar\affil{2}, and Tapio Schneider\affil{2}}

\affiliation{1}{Cecil H. and Ida M. Green Institute of Geophysics and Planetary Physics, Scripps Institution of Oceanography, University of California, San Diego, CA}

\affiliation{2}{California Institute of Technology, Pasadena, CA}


\begin{keypoints}
\item Ensemble Kalman methods effectively calibrate model parameters from noisy, time-averaged data
\item The numerical tests illustrate the robustness, applicability, and limitations of various versions of ensemble Kalman methods
\item No single method wins all ``races," but two methods stand out because they robustly estimate parameters in the absence of informative priors

\end{keypoints}


\begin{abstract}
Ensemble Kalman methods were initially developed to solve nonlinear data assimilation problems in oceanography but are now popular in applications far beyond their original use cases. 
Of particular interest is climate model calibration. 
As hybrid physics and machine-learning models advance, the number of parameters and complexity of parameterizations in climate models will continue to grow. To fully realize these advances, we must move from laborious hand-tuning to calibration-driven model development in rapid iteration cycles.
Thus, robust calibration of these parameters plays an increasingly important role. 
We focus on learning climate model parameters by minimizing the misfit between modeled and observed climate statistics in an idealized setting.
Ensemble Kalman methods are a natural choice for this problem because they are derivative-free, scalable to high dimensions, and robust to noise caused by statistical observations. 
Given the many variants of ensemble methods proposed, an important question is: \emph{Which ensemble Kalman method should be used for climate model calibration}? 
To answer this question, we perform systematic numerical experiments to explore the relative computational efficiencies of several ensemble Kalman methods.
The numerical experiments involve statistical observations of Lorenz-type models of increasing complexity, frequently used to represent simplified atmospheric systems, with some featuring neural network parameterizations.
For each test problem, several ensemble Kalman methods and a derivative-based method ``race'' to reach a specified accuracy, and we measure the computational cost required to achieve the desired accuracy. 
We investigate how prior information and the parameter or data dimensions define the computational costs of the various methods. 
\end{abstract}

\section*{Plain Language Summary}
The resolution of global climate models is too coarse to resolve many important but small-scale processes, such as clouds. Parameterization schemes are used to represent these unresolved atmospheric processes, including machine learned parameterizations.  These parameterizations contain many unknown parameters. A promising approach to improving models and their parameterizations is to calibrate them by matching model statistics (not trajectories) to observations. Ensemble Kalman methods are well suited for this task because they are derivative-free, scalable, and robust to the noisy data generated by aggregating climate statistics.  Since many variants exist, we investigate which ensemble method is most efficient for this calibration problem.  We set up systematic numerical experiments, ``racing'' several methods to see which reaches a specified accuracy at the lowest computational cost. Our findings reveal guidelines for the efficient use of ensemble methods for learning unknowns within climate models.


\section{Introduction}
Global climate models (GCMs) are used for many prediction horizons, including seasonal-to-decadal and beyond this.
To make climate projections, GCMs evolve a coarsely discretized state of the Earth system, often over several decades, and statistical projections are inferred from these trajectories. 
GCMs have two components: dynamics, defined by the laws of motion resolvable on the discretization grid-scale, and parameterizations, representing the effect of unresolved subgrid processes, e.g., cloud microphysics, convection, and turbulence.  
Depending on the current understanding of the processes being represented, parameterizations may rely on specified functional forms, sets of auxiliary differential equations, or they may need to be learned, e.g., with machine-learning (ML). 
To learn unknown parameters or functionals, GCMs can be calibrated to observations of Earth's climate \cite{SL17, SB23}. This shift enables modelers to move away from laborious hand-tuning and toward calibration-driven model development in rapid iteration cycles.

Calibration of GCM parameterizations by minimizing the mismatch between the time evolution (trajectory) of the model and the data is challenging because chaos drives an exponential divergence of trajectories.
The effects of chaos begin to dominate the misfit to data at relatively short lead-times of about two weeks. 
Even sophisticated techniques that calibrate states and parameters jointly with separate ensembles have seen limited success (e.g., \citeA{SJ15, RP13} and references therein).

A more recent and promising approach to GCM calibration involves minimizing a loss defined by the misfit of modeled and observed climate statistics \cite{SL24, SL17}.
A statistics-based loss function has the advantage that initial conditions are largely irrelevant (forgotten) and need not be computed; there is no chaotic divergence of statistical quantities (for ergodic systems), and the loss becomes primarily a function of the parameters. 
Learning a general physical constant independently from the initial conditions ensures the parameter's value will not strongly depend on them (this does not mean initial conditions are not important when making future predictions).
The disadvantage of using statistical quantities is that the chaotic variability of the climate system instead induces noise in the loss function \cite{DG21, CG21}. 
The noisy loss landscape and the lack of adjoints or automatic differentiation available for GCMs render the minimization of the loss via traditional, derivative-based optimization difficult. 
The noise level of statistical aggregates decreases slowly with accumulation time, and so even if gradients were available, a sufficient noise reduction for successful application of derivative-based approaches requires prohibitively long integration times \cite{CG21}. To enable the rapid iteration cycles required for modern parameterization development, the computational efficiency of these algorithms is paramount, especially given the large expense of GCM integrations.

A further challenge is that the GCM calibration problem is high-dimensional: For recent ML parameterizations, the unknowns are weights and biases of a (deep) neural network and could number hundreds, thousands, or many more.
Therefore, it is crucial to determine which optimization algorithms can tackle this high-dimensional and computationally demanding task efficiently, and without sacrificing accuracy.

One candidate class of algorithms with potential for GCM calibration goes by the name \emph{iterative ensemble Kalman methods}, which aim to approximate a Bayesian posterior distribution through an ensemble. 
The first ensemble Kalman methods were originally developed for solving nonlinear data assimilation problems (state estimation) in oceanography, with the most common being the ensemble Kalman filter (EnKF, \citeA{E94,E09}).
Over the years, many iterative versions of the EnKF have been created and applied for solving nonlinear inverse problems (parameter estimation). 
Examples of iterative ensemble methods include randomized maximum likelihood (EnRML, \citeA{GO07}), iterative ensemble smoothers\slash filters (IES\slash IEKF, \citeA{CO12, CO13}), the ensemble smoother with multiple data assimilation (ESMDA, \citeA{ER13, ER12}), and the iterative ensemble Kalman smoother (IEnKS, \citeA{BS14}).
Connections between various iterative ensemble methods were reported by \citeA{E18}.
Iterative ensemble methods are commonly used to match instances of sequential time-series data in reservoir engineering, numerical weather prediction (NWP), and physical oceanography, to name a few applications; however, they have not been thoroughly tested on noisy loss functions defined by statistical data.

Another class of ensemble-based algorithms goes by the name of \emph{ensemble Kalman inversion} (EKI, \citeA{IL13}).  
These algorithms are designed to optimize a loss function by collapsing an ensemble onto the minimizer of the loss.
The collapse of the ensemble distinguishes EKI from the iterative ensemble Kalman methods described above; it renders EKI methods unsuitable for uncertainty quantification.  
Variants of EKI algorithms include Tikhonov regularized EKI (TEKI, \citeA{CS20}), ensemble transform Kalman inversion (ETKI, \citeA{HH22}), and unscented Kalman inversion (UKI, \citeA{HS22}). 
Examples of EKI handling the noisy loss functions we expect to encounter during GCM calibration are shown in \citeA{CG21}, and some variants have already been applied successfully to climate science problems \cite{BD22, SS20, LC22, Christopoulos24a, King24a, Deck26a}.
Many EKI variants are also easily accessible in the Julia toolbox \verb|EnsembleKalmanProcesses.jl| \cite{DLG22}.

Ensemble Kalman methods are not restricted to Gaussian priors and likelihoods, or to quadratic loss functions.
These methods make use of ensemble Kalman updates that are derived from linear and Gaussian problems (Gaussian priors and likelihoods), but the overall result is \emph{not} a Gaussian or quadratic approximation. An EnKF ensemble, for example, in general does not have a Gaussian distribution, and EKI makes use of local statistical linearization in much the same way as Gauss-Newton methods use iterative, local linearization for the solution of nonlinear problems.
Our numerical experiments below demonstrate that EKI can successfully minimize complex loss landscapes that need not be quadratic. 

Given the many variants of EKI and iterative ensemble methods, an important question one should ask is: 
\emph{Which ensemble method should I use to estimate parameters from time-averaged data and chaotic dynamics}?  
In this paper, we contribute to finding an answer to this question by determining the relative computational efficiencies of several ensemble methods in a set of systematic and idealized prototype experiments. 
We focus on ensemble-based optimization and test diverse variants of EKI (TEKI, ETKI, UKI),
but we also consider IEKF as a representative of an iterative ensemble method.
Other iterative ensemble methods, e.g., ESMDA, require more tuning, e.g., for the number of iterations, which makes systematic numerical experiments difficult to perform.
We note future work should look closer at iterative ensemble methods, particle samplers (such as the ensemble Kalman sampler \cite{GH20,GN20}), particle-based accelerations (such as Calibrate, Emulate, Sample \cite{CG21,DG21}), and other ensemble-based Bayesian calibration approaches for climate science problems \cite{WG13, SF18, CB20} as parametric uncertainty quantification is important for predictions and projections with a GCM.


For each numerical experiment, we define chaotic dynamics using different Lorenz-type models \cite{L63,L95} and then use ensemble methods to minimize a loss function defined by the misfit between modeled and observed statistics.
Each ensemble method must reach a specified accuracy, and we evaluate performance of this race by the number of forward model runs needed to reach that accuracy (since the forward model runs, rather than the ensemble updates, dominate the computational cost in our applications of interest).
For each ensemble method, we first determine an optimal ensemble size to maximize computational efficiency (with the exception of UKI, for which the ensemble size is a function of the number of unknowns).
To highlight the difficulties derivative-based optimization encounters with the test problems, we also show for each problem that the Levenberg-Marquardt (LM) method, a nonlinear least squares solver \cite{L44, M63, NW06}, fails to converge to a useful solution, because the solver is too easily trapped in one of many local minima.
For context, \citeA{MW09} also compared various derivative-free optimization algorithms (before the advent of the methods discussed here), but their work was not focused on inverting statistical observations and the comparison was done assuming a fixed computational budget (and consequently varying accuracy). 
The computational expense and importance of GCM calibration, however, more naturally leads to finding the most efficient algorithm for a fixed accuracy.

Our systematic experiments address several important aspects of ensemble-based optimization, relevant to GCM calibration. 
\begin{enumerate}
    \item 
    We increase the complexity and dimensionality of the test problems, so our experiments can indicate which ensemble methods may have the greatest potential to scale to high-dimensional GCM calibration.
    \item
    We investigate the role the prior plays in ensemble methods.
    Intuitively, a good, informative prior should accelerate the convergence and, hence decrease the computational cost of ensemble methods.
    Reliable priors, however, are not always available.
    Our experiments highlight which ensemble methods are robust to the lack of reliable prior knowledge.
    \item One of our test problems features a neural network parameterization, this reflects an important recent direction in the context of GCM calibration where ML parameterizations are proposed to represent subgrid processes. 
    The numerical examples we present here are a first step towards understanding issues that may arise when using neural network parameterizations for GCM calibration.
    \item 
    We run many repeated test problems under random re-initializations, and provide quantitative statements regarding the robustness of our results and stability of the algorithms.
\end{enumerate}

The rest of the paper is organized as follows.  
In Section~\ref{sec:EKI} we briefly review the ensemble methods we compare.  
Section~\ref{sec:three} describes the systematic numerical experiments: Definitions of forward models (including dynamics, parameterizations, and priors), statistical observations and associated error models, iteration stopping criteria, and optimal ensemble sizes.  
We report and discuss the outcome of the numerical experiments in Section~\ref{sec:results} and conclude the paper with a summary of our findings in Section~\ref{sec:conclusions}.

\section{Variants of iterative ensemble methods}
\label{sec:EKI}

Iterative ensemble methods for parameter estimation originated from the ensemble Kalman filter (EnKF, \citeA{E09, E94}), which has been vastly successful in numerical weather prediction (NWP), physical oceanography, and reservoir engineering.
To stick to the example of NWP, the EnKF updates a forecast ensemble, generated by a numerical model, with atmospheric data. 
The mean of the updated ensemble represents the predicted state of the atmosphere and the ensemble itself represents variations in the atmospheric state that are compatible with the available data.  
The EnKF iterates this process, and each cycle of the EnKF assimilates the most recent observations.  

Iterative ensemble methods use the EnKF formalism iteratively on \emph{the same} data set.
Since an EnKF update (i.e., an analysis step) reduces variance, the spread of the ensemble in iterative methods shrinks after each iteration. 
In ensemble Kalman inversion (EKI), the repeated use of the same data causes the ensemble to ultimately collapse onto the minimizer of a loss function.
Other iterative ensemble methods, e.g., IEKF or ESMDA, prevent the collapse of the ensemble, e.g., by inflating the observation error covariances appropriately.
As a result, the ensemble in those iterative ensemble methods does not collapse and is approximately distributed according to a Bayesian posterior distribution \cite{ER13,CO13,CHSV}.

\subsection{The EKI loss function and notation}
The loss function targeted by EKI is
\begin{linenomath*}
\begin{equation}
\mathcal{L}\left(u\right) = \frac{1}{2}\biggl\|{R_d^{-\frac{1}{2}}\left(d - \mathcal{F}\left(u\right)\right)}\biggr\|^2 + 
				\frac{1}{2}\biggl\|B^{-\frac{1}{2}}\left(m - u\right)\biggr\|^2,
\label{eq:loss}
\end{equation}
\end{linenomath*}
where $\|\cdot \|$ denotes the two-norm, $\mathcal{F}(\cdot)$ is the nonlinear forward model mapping parameters $u$ to the data $d$, and $R_d$ is the data error covariance matrix. 
The first term in~\eqref{eq:loss} describes how well the model fits the data, and the second term regularizes the optimizer to be close to the vector $m$, with ``closeness'' being specified by the symmetric positive definite matrix $B$.
Within a Bayesian context, this loss forms the negative log-likelihood, with the first term attributed to a Gaussian likelihood $p_l(d\vert u)$ and the second term attributed to a Gaussian prior $p_0(u)$ with mean $m$ and covariance matrix $B$.
The posterior distribution defined by the likelihood and prior, $p(u\vert d) = p_0(u)p_l(d\vert u)/p(d)$, is proportional to $\exp(-\mathcal{L}(u))$.

The loss function~\eqref{eq:loss} can be rewritten as 
\begin{linenomath*}
\begin{equation}
\mathcal{L}\left(u\right) = \frac{1}{2}\left\|R^{-\frac{1}{2}}\left(y - \mathcal{G}\left(u\right)\right)\right\|^2,
\label{eq:recast-loss}
\end{equation}
\end{linenomath*}
where
\begin{linenomath*}
\begin{equation*}
            R = \begin{pmatrix} 
				R_d & 0 \\
				0 & B 
			\end{pmatrix}, \qquad 
            y = \begin{pmatrix}
			d \\
			m
			\end{pmatrix}, \qquad 
            \mathcal{G}(u) = \begin{pmatrix}
			\mathcal{F}(u) \\
			u
			\end{pmatrix}.
\end{equation*}
\end{linenomath*}
Here, $\mathcal{G}(\cdot)$ augments the model with the identity map, $y$ is an extended observation (data and prior mean), and the extended observation error covariance is $R$ (accounting for observation error covariances $R_d$ and prior covariances $B$). 
The variants of EKI we describe below can all be compactly described with reference to the loss~\eqref{eq:recast-loss}; however, IEKF distinguishes the prior covariance $B$ from the data error covariance $R_d$ and, hence, is more conveniently described with reference to~\eqref{eq:loss}.

We use subscripts for the iteration and superscripts for the ensemble members, i.e.,
$u_j^k$ is the $k$th ensemble member at iteration $j$.
The various ensemble methods make use of normalized perturbation matrices
\begin{linenomath*}
\begin{align}
U_j & = \frac{1}{\sqrt{n_{\text{e}} - 1}}\left(u_j^1 - \bar{u}_j \; \; u_j^2 - \bar{u}_j \; \; \cdots \; \; u_j^{n_{\text{e}}} - \bar{u}_j\right), & \bar{u}_j = \frac{1}{n_{\text{e}}}\sum_{k = 1}^{n_{\text{e}}} u_j^k,\\
G_j & = \frac{1}{\sqrt{n_{\text{e}} - 1}}\left(g_j^1 - \bar{g}_j \; \; g_j^2 - \bar{g}_j \; \; \cdots \; \; g_j^{n_{\text{e}}} - \bar{g}_j\right), & \bar{g}_j = \frac{1}{n_{\text{e}}}\sum_{k = 1}^{n_{\text{e}}} g_j^k,
\end{align}
\end{linenomath*}
where $g_j^k = \mathcal{G}(u_j^k)$.
Empirical covariance and cross-covariance matrices can be defined via the perturbation matrices as
\begin{linenomath*}
\begin{align}
C_j^{uu}  = U_j U_j^\mathsf{T},\quad 
C_j^{gg}  = G_j G_j^\mathsf{T},\quad
C_j^{ug}  = U_j G_j^\mathsf{T},
\label{eq:covs}
\end{align}
\end{linenomath*}
which describe covariances between the unknowns ($C_j^{uu}$), between the predicted observation ($C_j^{gg}$), and cross-covariances between the unknowns and predicted observation ($C_j^{ug}$); here and below, superscript $\mathsf{T}$ denotes the transpose.

\subsection{Tikhonov regularized EKI}
\label{sec:TEKI}
Tikhonov regularized EKI (TEKI, \citeA{CS20}) iterates the update 
\begin{linenomath*}
\begin{equation}
u_{j + 1}^{k} = u_j^k + \alpha \left\{ C_{j}^{ug}\left(C_{j}^{gg} + R \right)^{-1}\left(y - \left(g_j^k + \eta_j^k\right)\right) \right\},
\label{eq:TEKI}
\end{equation}
\end{linenomath*}
where $\eta_j^k$ is a random sample from a Gaussian distribution with mean zero and covariance matrix~$R$ and where $\alpha$ is a step size (which we typically set to $\alpha=1$).
Note that~\eqref{eq:TEKI} basically amounts to repeated EnKF updates using the same data $y$. 
The spread of the ensemble is reduced at each iteration, leading to TEKI ensemble collapse onto the minimizer of the loss~\eqref{eq:recast-loss} \cite{SS16, SS18, CT19}. 
Though iterations of TEKI and EKI are identical, the first paper on EKI \cite{IL13} did not consider regularization or, equivalently, prior distributions (i.e., the loss function~\eqref{eq:loss} has no second, quadratic term).

\subsection{Ensemble transform Kalman inversion}
\label{sec:ETKI}

Ensemble transform Kalman inversion (ETKI, \citeA{HH22}) is based on the ensemble transform Kalman filter (ETKF, \citeA{BE01,TABHW03}).  
During an iteration, ETKI updates the ensemble mean $\bar{u}_j$ and the ensemble perturbations $U_j$ separately.
The update equation for the ensemble mean is 
\begin{linenomath*}
\begin{equation}
    \bar{u}_{j + 1} = \bar{u}_j + U_j\left(I + G_j^\mathsf{T} R^{-1} G_j\right)^{-1} G_j^\mathsf{T} R^{-1}  \left(y - \bar{g}_j \right),
\label{eq:ETKImean}
\end{equation}
\end{linenomath*}
and the ensemble perturbations are updated by
\begin{linenomath*}
\begin{equation}
        U_{j +1} = U_jT, \qquad T =\left(I + G_j^\mathsf{T} R^{-1} G_j\right)^{-\frac{1}{2}},
\label{eq:transformation}
\end{equation}
\end{linenomath*}
where $T$ is a \emph{symmetric} $n_{\text{e}}\times n_{\text{e}}$ ensemble transform matrix.
The ensemble for the next iteration is constructed by adding the updated perturbations (columns of the matrix $U_{j+1}$) to the updated ensemble mean $\bar{u}_{j+1}$:
\begin{linenomath*}
\begin{equation}
    u_{j + 1}^{k} = \bar{u}_{j + 1} + [U_{j +1}]_k\sqrt{n_{\text{e}} - 1};
\label{eq:ETKI}
\end{equation}
\end{linenomath*}
here $[U_{j +1}]_k$ is the $k$th column of $U_{j+1}$.
To ensure that the ensemble mean is $\bar{u}_{j + 1}$, one must chose a symmetric square root in \eqref{eq:transformation}.
ETKI is numerically efficient for the common situation where the ensemble size $n_{\text{e}}$ is smaller than the number of observations or the number of unknowns, because the method performs all computations in the \emph{ensemble space}, so linear solves and matrix square roots are done with small matrices of size $n_{\text{e}}\times n_{\text{e}}$.
ETKI is most efficient if one can efficiently compute the matrix $R^{-1}G_j$, e.g., when $R$ is low-rank, diagonal, or sparse.
We note that the ETKI and TEKI updates are asymptotically equivalent, but for a finite ensemble size the ensembles are different \cite{AS23} --for example, the TEKI ensemble depends on the random perturbations $\eta_j^k$ in~\eqref{eq:TEKI}.

\subsection{Unscented Kalman inversion}
\label{sec:UKI}
Unscented Kalman inversion (UKI, \citeA{HS22}) is based on the unscented Kalman filter (UKF, \citeA{WV00}) and the unscented transform \cite{JU97}.  
The basic idea is to use the unscented transform to define the ensemble by a quadrature stencil known as \emph{sigma points}.
There are various ways to define sigma points, but we follow \citeA{HS22} and define the ensemble by the sigma points:
\begin{linenomath*}
\begin{align}
    & u_j^1  = \bar{u}_j, \\
    & u_{j}^{k + 1}  = \bar{u}_j + a\sqrt{n_u}\left[\left(C_j^{uu}\right)^{\frac{1}{2}}\right]_k \quad \left(1 \leq k \leq n_u\right), \\
    & u_j^{k + 1 + n_u}  = \bar{u}_j - a\sqrt{n_u}\left[\left(C_j^{uu}\right)^{\frac{1}{2}}\right]_k \quad \left(1 \leq k \leq n_u\right).
\end{align}
\end{linenomath*}
Here, $a$ is a hyper-parameter that we set to one and $[(C_j^{uu})^{\frac{1}{2}}]_k$ is the $k$th column of the Cholesky factor of $C_j^{uu}$;  
for the first iteration, $C_0^{uu} = B$ and $\bar{u}_0 = m$.   
The ensemble mean and covariance matrix are updated using
\begin{linenomath*}
\begin{align}
    \bar{u}_{j+1} & = \bar{u}_j + C_j^{ug}\left(C_j^{gg} + a^2R\right)^{-1}\left(y - \bar{g}_j\right), \\
    C_{j + 1}^{uu} & = C_j^{uu} - C_j^{ug}\left(C_j^{gg} + a^2R\right)^{-1} \left( \frac{1}{a^2}C_j^{ug} \right)^\mathsf{T},
\end{align}
\end{linenomath*}
and the UKI ensemble $u_{j+1}^k$ is generated by computing new sigma points from $\bar{u}_{j + 1}$ and $C_{j+1}^{uu}$.
Importantly, the ensemble size of UKI is fixed:
\begin{linenomath*}
\begin{equation}
\label{eq:UKIEnsembleSize}
    n_{\text{e}} = 2n_u+1.
\end{equation}
\end{linenomath*}
This can be efficient when the dimension $n_u$ of the parameter space is small, but becomes problematic when $n_u$ is large (as expected in GCM calibration), because each ensemble member requires a forward model evaluation, which can be computationally expensive.

\subsection{The iterative ensemble Kalman filter}
\label{sec:IEKF}
The iterative ensemble Kalman filter (IEKF) was first proposed by \citeA{CO13} and later studied by \citeA{CC21} (who refer to it as IEKF with statistical linearization). 
IEKF is derived from the Gauss-Newton update
and uses an ensemble approximation for the Jacobian of the forward model $\mathcal{F}(\cdot)$, including the normalized perturbation matrix
\begin{linenomath*}
\begin{align}
F_j & = \frac{1}{\sqrt{n_{\text{e}} - 1}}\left(f_j^1 - \bar{f}_j \; \; f_j^2 - \bar{f}_j \; \; \cdots \; \; f_j^{n_{\text{e}}} - \bar{f}_j\right), & \bar{f}_j = \frac{1}{n_{\text{e}}}\sum_{k = 1}^{n_{\text{e}}} f_j^k,
\end{align}
\end{linenomath*}
where $f_j^k = \mathcal{F}(u_j^k)$. 
Specifically, at iteration $j$, the Jacobian approximation is
\begin{linenomath*}
\begin{equation}
    J_j = F_j U_j^{\dagger},
    \label{eq:psuedo-inverse}
\end{equation}
\end{linenomath*}
where the superscript $\dagger$ denotes a pseudo-inverse.
With this Jacobian approximation, the IEKF iteration is
\begin{linenomath*}
\begin{equation}
u_{j + 1}^k = u_j^k + \alpha \left\{ K_j\left(d - \left(f_j^k + \varepsilon_j^k\right)\right) + \left(I - K_jJ_j\right)\left(m - \left(u_j^k + \zeta_j^k\right)\right) \right\}, 
\label{IEKF}
\end{equation}
\end{linenomath*}
where
\begin{linenomath*}
\begin{align}
K_j &= B J_j^\mathsf{T} \left(J_j B J_j^\mathsf{T} + R_d\right)^{-1}, \\
\varepsilon_j^k & \sim \mathcal{N}(0,2\alpha^{-1}R_d),\\
\zeta_j^k & \sim \mathcal{N}(0,2\alpha^{-1}B).
\label{KH}
\end{align}
\end{linenomath*}
We note that there are slight differences between IEKF of \citeA{CO13} and \citeA{CC21} regarding the scaling of the random perturbations $\varepsilon_j^k$ and $\zeta_j^k$.
In the numerical experiments below, we use the implementation described by \citeA{CC21}, but similar results are obtained when we use the implementation originally proposed by \citeA{CO13}.

\subsection{Derivative-based optimization with Levenberg-Marquardt}

The Levenberg-Marquardt (LM) method is a classical, derivative-based nonlinear least squares method \cite{L44, M63, NW06}.
To derive the LM update equations, we write the loss function~\eqref{eq:recast-loss} as the two-norm of a residual $r(u)$,
\begin{linenomath*}
\begin{equation}
    \mathcal{L}(u) = \frac{1}{2}\left\|r(u) \right\|^2, \quad r(u) = R^{-\frac{1}{2}}(y-\mathcal{G}(u)).
\end{equation}
\end{linenomath*}
In LM, the residual $r(u)$ is approximated by its linearization around the current iterate $u_j$ and an increment $\Delta u_j = u_{j+1}-u_j$ is defined by the solution of the regularized linear least squares problem
\begin{linenomath*}
\begin{equation}
\label{eq:LM_LS}
    \min_{\Delta u_j} \frac{1}{2}\left\| r(u_j) + \frac{\partial r}{\partial u}\vert_{u_j}\Delta u_j\right\|^2+\frac{1}{ 2\lambda}\left\| \Delta u_j\right\|^2,
\end{equation}
\end{linenomath*}
where $\lambda>0$ is a damping factor adjusted during the iterations and where $ \frac{\partial r}{\partial u}\vert_{u_j}$ is the Jacobian of the residual, evaluated at the current iteration $u_j$.
Writing the Jacobian of $r(u)$ in terms of the Jacobian of the generalized model $\mathcal{G}$
\begin{linenomath*}
\begin{equation}
    \frac{\partial r}{\partial u} = -R^{-\frac{1}{2}}\frac{\partial \mathcal{G}}{\partial u}
\end{equation}
\end{linenomath*}
and substitution into~\eqref{eq:LM_LS} gives
\begin{linenomath*}
\begin{equation}
    \min_{\Delta u_j} \frac{1}{2}\left\| R^{-\frac{1}{2}}\left((y-\mathcal{G}(u_j)) -D_j\Delta u_j \right)\right\|^2+\frac{1}{2 \lambda}\left\| \Delta u_j\right\|^2,
\end{equation}
\end{linenomath*}
where we introduced the shorthand notation $D_j = \partial\mathcal{G}/\partial u \vert_{u_j}$ for the Jacobian of the generalized model, evaluated at the current iterate $u_j$.
The solution of this least squares problem is
\begin{linenomath*}
\begin{equation}
    \Delta u_j^* = \lambda D^\mathsf{T}(\lambda DD^\mathsf{T}+R)^{-1}(y-\mathcal{G}(u_j)),
\end{equation}
\end{linenomath*}
so that the LM update equation becomes
\begin{linenomath*}
\begin{equation}
    u_{j+1} = u_j + \Delta u_j^* = u_j +\lambda D^\mathsf{T}(\lambda DD^\mathsf{T}+R)^{-1}(y-\mathcal{G}(u_j)).
\end{equation}
\end{linenomath*}
We use LM within Python and SciPy's \verb|least_squares| algorithm, which in turn is a wrapper around the \verb|MINPACK| Fortran library.
In our numerical experiments, LM is a representative of derivative-based optimization methods, and we compare LM to ensemble-based methods in all numerical experiments.

The main purpose of this comparison is to demonstrate that the loss functions in our idealized test problems indeed pose problems for classical optimization.
We will see that the losses are so noisy that LM (or other derivative-based optimizers) are trapped in local minima. The ensemble methods can overcome this issue and optimize the noisy loss functions.
In the future, evaluating the efficiency of modern derivative-based optimization (e.g., large step sizes \cite{Wangetal2026} or stochastic gradient descent \cite{Goodfellow2016}) would be interesting, but here we focus on testing the closest-related derivative-based method, LM, to the ensemble methods.

\section{Experiment design}
\label{sec:three}

Our numerical experiments are designed to explore the applicability of ensemble methods for GCM calibration.
The basic idea is to define a forward model using a chaotic dynamical system that depends on model parameters or parameterizations. 
We apply various ensemble methods (TEKI, ETKI, UKI, IEKF) to estimate model parameters or parameterizations from time-averaged data. 
We then measure the computational cost each ensemble method requires to find a solution with a specified accuracy.

We define the dynamics, model parameters, and parameterizations, the time averaged data and associated error models in Section~\ref{sec:time}.
Prior distributions are described in Section~\ref{sec:prior}.
The only tunable parameter of the ensemble methods is the ensemble size $n_{\text{e}}$; we compute and use an optimal ensemble size that maximizes computational efficiency (Section~\ref{sec:opt}).
Section~\ref{sec:RMSE} defines the exit criteria for optimization iterations that ensures that each ensemble method finds parameters or parameterizations that lead to similar model-data misfits, or target accuracies.
In our experiments, we assess the computational cost of EKI variants by the number of forward model runs needed to reach the target accuracy.
This is a reasonable proxy for the computational cost of an ensemble method, because GCM simulations are computationally much more expensive than the linear algebra required for EKI ensemble updates.

Systematic numerical experiments like these are currently out of reach with GCMs:
Our sampling strategy for computing the optimal ensemble size is too expensive for GCMs.
To reduce the computational burden of the numerical experiments, we use systems of chaotic ordinary differential equations (Lorenz-type problems, see Section~\ref{sec:time}) instead of GCMs. 
While this is a drastic simplification, the numerical experiments with simplified dynamics can systematically investigate some important aspects of ensemble-based methods, relevant to GCM calibration. 
We repeat the experiments under randomized initializations to ensure that our conclusions are robust to choices of the initial ensemble.
We thus investigate which ensemble methods have the greatest \emph{potential} to scale to GCM 
calibration, and what role informative versus uninformative priors play in this context.
One of our test problems includes a neural network parameterization; the implied loss function is not only noisy, but also non-convex
(in practice, we may encounter non-convexity in situations that do not involve a neural network).
Our test problems highlight how changes to the loss landscape can impact the ensemble methods' ability to minimize the loss function.

\subsection{Forward models, time-averaged data, and error models}
\label{sec:time}

\begin{figure}[tb]
    \centering
    \includegraphics[width=0.99\linewidth]{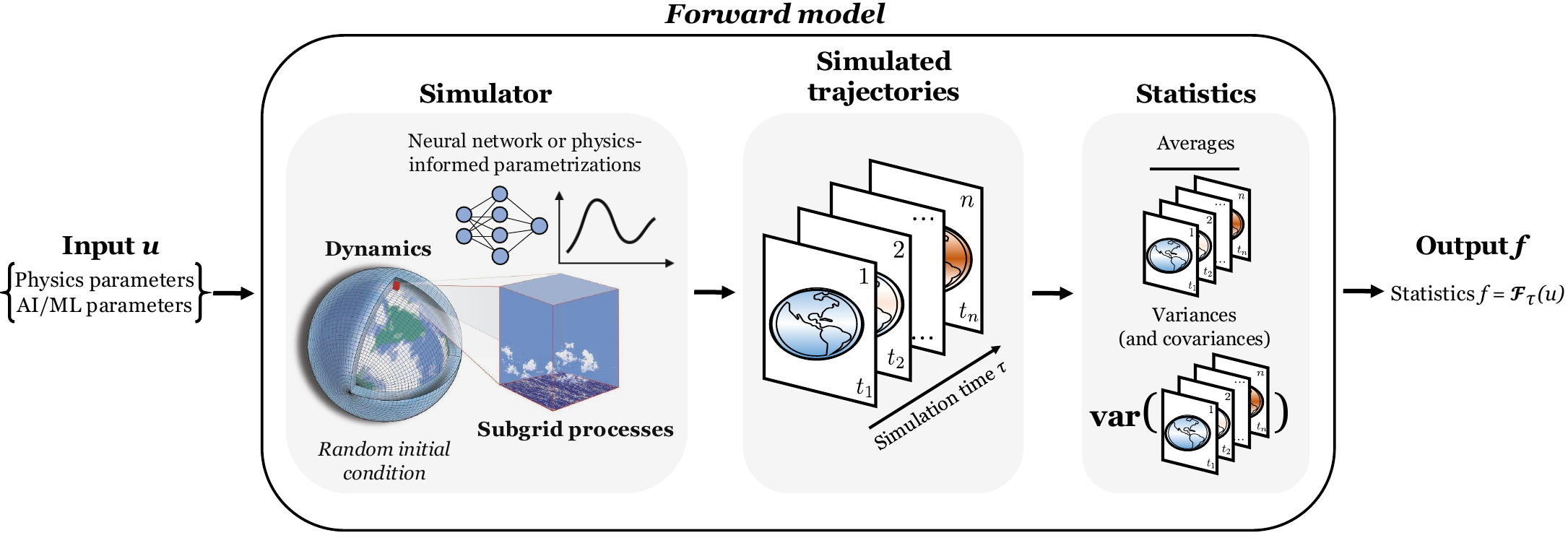}
    \caption{
    Flow chart of a forward model $\mathcal{F}_\tau(\cdot)$ whose outputs are statistics of the system state.
    For a given set of parameters $u$, the forward model first simulates the dynamics for a time period $\tau$, starting from a random initial condition. The statistics $f=\mathcal{F}_\tau(u)$ are subsequently computed from the simulation output after a spin-up period. (Image of dynamics and subgrid processes simulation is from \citeA{ST17}.)}
    \label{fig:forward_problem}
\end{figure}

The forward model $\mathcal{F}_\tau(\cdot)$ is the process of simulating a dynamical model for $\tau$ model time units (all models we consider are non-dimensional) and subsequently computing statistics of the simulation output.
The forward model for time-averaged data is illustrated by a flowchart in Figure~\ref{fig:forward_problem},
and explained in detail below.

\subsubsection{Dynamics and parameterizations}
In our experiments, the dynamical models are Lorenz-type, chaotic, ordinary differential equations (ODE), which we discretize with a fourth-order Runge-Kutta scheme.
Specifically, we use the Lorenz '63 model (L'63, \citeA{L63})
\begin{linenomath*}
\begin{equation}
    \frac{\text{d}z_1}{\text{d}t} = \sigma\left(z_2 - z_1 \right), \quad \frac{\text{d}z_2}{\text{d}t} = \rho z_1 - z_2 - z_1z_3, \quad \frac{\text{d}z_3}{\text{d}t} = z_1z_2 - \beta z_3,
\label{eq:L63}
\end{equation}
\end{linenomath*}
where $\sigma,\rho$ and $\beta$ are model parameters.
In the numerical experiments in Section~\ref{sec:L63}, we estimate $\rho$ and $\beta$, but fix $\sigma=10$.
We also consider the Lorenz '96 model (L'96, \citeA{L95}), defined by the ODE
\begin{linenomath*}
\begin{equation}
\frac{\text{d}z_l}{\text{d}t} = z_{l-1}\left(z_{l+1} - z_{l-2}\right) - z_l + \phi_l, \quad l = 1, \dots, n_z, 
\end{equation}
\end{linenomath*}
on a periodic domain, where $\phi_l$ is a forcing.
We use the L'96 model with two different forcings:
\begin{enumerate}
    \item 
    The classical constant forcing
    \begin{linenomath*}
    \begin{equation}
        \phi_l=\phi = 8, \quad l=1,\dots,n_z.
    \end{equation}
    \end{linenomath*}
    \item 
    A spatially varying forcing \cite{VM24},
    \begin{linenomath*}
    \begin{equation}
    \label{eq:SpatiallyVaryingL96Forcing}
        \phi_l =  8 + 6\text{ sin}\left(\frac{4 \pi}{n_z} l \right), \quad l=1,\dots,n_z.
    \end{equation}
    \end{linenomath*}
\end{enumerate}
In the case of a constant forcing, the ensemble methods estimate $\phi$.
To emulate the effects of parameterizations in GCMs, we consider different parameterizations of the spatially varying forcing: a grid-based parameterization, where we define each $\phi_l$ separately (as is the case in the actual dynamics), and a neural network parameterization.
The latter setup mimics, to some extent, climate models with subgrid processes parameterized by neural networks.
Note that in the grid-based configuration, the ensemble methods estimate $\phi_l$ on the grid, but in the neural network parameterization, the ensemble methods estimate the weights and biases of a neural network that parameterizes the forcing as a function of the spatial index $l$.
The different parameterizations of the forcing result in different numbers of unknowns across the test problems; we further vary the state dimension $n_z$ to design test problems with different complexities.
The various configurations of the dynamics and parameterizations are summarized in Table~\ref{tab:Dynamics}.
\begin{table}
    \centering
    \begin{tabular}{ccccc}
         Dynamics  & Parameterization ($u$)  & \# of unknowns ($n_u$) & State dimension ($n_z$) \\\hline
         L'63&  Model parameters $\rho,\beta$       &   2  &   3 \\
         L'96&  Model parameter $\phi$              &   1  &  40 \\
         L'96&  Grid-based $\phi_l$                 &  40  &  40 \\
         L'96&  Neural net parameterized $\phi_l$   &  61  & 100 \\
    \end{tabular}
    \caption{Summary of the dynamical models and parameterizations in our numerical experiments.
    The dynamics are Lorenz models (L'63 or L'96). 
    The parameterizations determine $n_u$ parameters --  either model parameters or parameterizations.
    The number of unknowns is, in general, different from the number of state variables of the dynamics or the number of statistical observations.
    }
    \label{tab:Dynamics}
\end{table}

\subsubsection{Time-averaged data and associated error models}
\label{sec:DataAndErrorModel}
We consider estimation of model parameters or parameterizations from time-averaged data.
The forward model $\mathcal{F}_\tau(\cdot)$ thus amounts to simulating the dynamics for a time interval $\tau$, starting from a randomized initial condition $z(0)$, followed by evaluating first- and second-order moments of the system state after a ``spin-up period'' \cite{DG21, CG21}.
The randomized initial condition implies a spin-up, during which the influence of a randomly chosen initial condition is ``forgotten'' by the chaotic dynamics.

For the L'63 dynamics, the statistics are the mean, variances, and covariances of the three state variables \cite{CG21}:
\begin{linenomath*}
\begin{equation}
    \mathcal{F}_\tau^\text{L'63}(u)= (\bar{z}_1, \bar{z}_2, \bar{z}_3, \text{var}(z_1), \text{var}(z_2), \text{var}(z_3), \text{cov}(z_1,z_2), \text{cov}(z_1,z_3), \text{cov}(z_2,z_3)).
    \label{eq:L63_data}
\end{equation}
\end{linenomath*}
For the L'96 dynamics, the statistics are the mean and standard deviations of the $n_z$ state variables:
\begin{linenomath*}
\begin{equation}
    \mathcal{F}_\tau^\text{L'96}(u)= (\bar{z}_1, \cdots, \bar{z}_{n_z}, \sqrt{\text{var}(z_1)}, \cdots, \sqrt{\text{var}(z_{n_z})}).
    \label{eq:L96_data}
\end{equation}
\end{linenomath*}
For each controlled, systematic numerical experiment, we can generate synthetic data as follows: We apply the forward model $\mathcal{F}_\tau(\cdot)$ to a ``true'' parameter set $u_\text{true}$ and ``true'' initial condition $z_{\text{true}}(0)$ to obtain the statistical quantities $d=\mathcal{F}_\tau(u_\text{true}, z_{\text{true}}(0))$; the data.

An important characteristic of the forward model is that it is noisy, i.e., a fixed parameter $u$ with different initial conditions can result in different model outputs $f$.
The noise is caused by the randomized (unknown) initial condition and the short simulation time $\tau$.
Specifically, let $\mathcal{F}_\infty(\cdot)$ denote the statistics one would compute from an \emph{infinitely long} simulation (for any initial condition).
Then
\begin{linenomath*}
\begin{equation}
    \label{eq:InftyModelAndNoise}
    \mathcal{F}_{\tau}\left(u, z(0)\right) \approx \mathcal{F}_{\infty}\left(u\right) + \frac{1}{\sqrt{\tau}}\xi,
\end{equation}
\end{linenomath*}
where $\xi$ is normally distributed with mean zero \cite{CG21}.
In other words, the noise level is related to the duration of the simulation $\tau$.  The longer the time window, the less noisy are the statistics we compute;
a shorter time window leads to noisier statistics.  

We make use of~\eqref{eq:InftyModelAndNoise} to define the observation error covariance $R_d$ \cite{CG21}.
Recall that the data are statistical quantities computed from a simulation of duration $\tau$.
To compute the observation error covariance, we run a simulation of duration $n_{\text{s}}\cdot \tau$ and then window the simulation output to compute $n_{\text{s}}$ samples of statistical data associated with a simulation duration $\tau$.
The observation error covariance is then simply the sample covariance.
This strategy of computing $R_d$ naturally takes into account that the noise in the statistical data decreases with the simulation time $\tau$, but requires a rather long overall simulation time ($n_{\text{s}}\cdot \tau$), rendering this strategy infeasible for GCM calibration, though sampling-error estimators can help \cite{VM24, LW22}.
One can also consider porting techniques from data assimilation to estimating error covariances for statistical data   \cite{PT18,TA20}.

Table~\ref{tab:Data} summarizes the simulation time $\tau$, the spin-up, and the number of simulations we use for the construction of the observation error covariance matrix $R_d$ in the numerical experiments.
\begin{table}
    \centering
    \begin{tabular}{cccccc}
         Dynamics  & Parameterization ($u$)  & Sim. time ($\tau$) & \# of sims. ($n_{\text{s}}$) & Spin-up & \# of data \\\hline
         L'63&  Model param. $\rho,\beta$     &  10 &   36 & 30 &   9\\
         L'96&  Model param. $\phi$           &  10 &  800 &  4 &  80\\
         L'96&  Grid-based $\phi_i$           &  50 &  800 &  4 &  80\\
         L'96&  Neural Net                    &  50 & 2000 &  4 & 200\\
    \end{tabular}
    \caption{Summary of the specifications the forward models use in the numerical experiments.
    The dynamics are Lorenz models (L'63 or L'96). 
    The inversions recover different parameters, defined by different types of parameterizations.
    The inversions compute statistical quantities based on a simulation of the dynamics of duration $\tau$.
    The construction of the observation error covariance matrix $R_d$ is based in a simulation of duration $n_{\text{s}}\cdot \tau$.
    Statistical data are computed after a spin-up period during which the influence of unknown initial conditions is forgotten.
    }
    \label{tab:Data}
\end{table}
We use a default time step of $\Delta t= 0.01$ (for L63 and L96), but for the modified Lorenz problem with a neural network parameterization, we decrease the time step to $\Delta t= 0.005$, because the smaller time step stabilizes the optimizations.
Note that the simulation times $\tau$ are relatively short throughout our experiments, which renders the forward models and implied loss functions noisy (too noisy for derivative-based optimization, as we will demonstrate).
The number of simulations used to estimate the observation error covariance matrix ($n_{\text{s}}$) scales with the number of data $n_d$, to ensure accuracy in each setting.
Since we are computing statistics for each state variable, the number of data $n_d$ scales with the state dimension $n_z$.
For GCMs, such long simulations times may be overly expensive, but with our simplified setup, accurate modeling of errors due to chaotic internal variability is feasible.
In practice, model and measurement errors also arise, but these play no role in our perfect model experiments and we, for now, ignore structural model errors.

\subsubsection{Summary of the forward model: dynamics, parameterizations, statistical data and observation errors}
In summary, we have defined the forward models $\mathcal{F}_\tau(\cdot)$ that use different dynamics and model parameters or parameterizations.
The forward model maps model parameters\slash parameterizations to statistical data, and we described how to accurately model errors in these data arising from chaotic internal variability via a careful construction of the observation error covariance matrix $R_d$.
The synthetic data $d$ for each optimization experiment is generated by applying the forward model $\mathcal{F}_\tau(\cdot)$ to $u_{\text{true}}$ and $z_{\text{true}}(0)$.

We emphasize again that the statistical data are generated with an initial condition that is unknown to the ensemble methods we use to recover the true parameter set (the initial condition is randomized in the forward model).
The fact that the initial conditions need not be computed is an advantage of using statistical data.

\subsection{Prior distributions}
\label{sec:prior}
For the L'63 dynamics, we follow \citeA{CG21}, and define the prior as independent log-normal distributions with location and scale parameters corresponding to $m=(3.3, 1.2)$ and $B=\mathrm{Diagonal}(0.5^2, 0.15^2)$ (i.e., we use a Gaussian prior and move exponentiation to the forward map); this prior distribution yields positive distributions for $\rho$ and $\beta$ with means $(30.7,3.36)$, and with a (0.05,0.95) quantile range of $(11.9,61.6)$ for $\rho$ and $(2.59,4.25)$ for $\beta$.
The true parameter values are $\rho=28$ and $\beta=8/3$, lying in a region of high prior probability.

All experiments with L'96 dynamics feature Gaussian priors.  For the L'96 dynamics with constant forcing, the prior is centered close to the true forcing with prior mean $m = 10$ and standard deviation 4. As before, the true parameter value ($\phi_\text{true}=8$) lies in the region of high prior probability.
For L'96 with spatially variable forcing and a grid-based parameterization, the prior mean is $m=(8,\dots,8)$ 
and the elements $B_{ij}$ of the prior covariance are
\begin{linenomath*}
\begin{equation}
    B_{ij} = 9 \exp\left( -\frac{1}{2}\min \left(\vert i-j\vert, n_u - \vert i-j\vert \right)\right),
\end{equation}
\end{linenomath*}
with the periodic boundaries incorporated appropriately. 
Again, the true parameter values are well within the region of high prior probability.

The construction of the prior for the L'96 model with a neural network parameterization is more difficult
because there are no clear guidelines for how to construct informative prior distributions for weights and biases of a neural network.
Simply using a zero mean and moderate standard deviation leads to blow-up of all simulations.
We describe two differently structured prior distributions to evaluate the effects of informative and largely uninformative priors on calibration performance.
Recent work on estimating the uncertainties of NN weights and biases includes \citeA{VC25}; however, this approach assumes that sufficient training and testing data are available.

We first consider an uninformative prior which we construct following \cite{VM24} and which simply accounts for the prior knowledge that the forcing function should be smooth.
To generate the mean and covariance matrix of the uninformative Gaussian prior, we generate random smooth functions by drawing samples from a periodic Gaussian process with a Gaussian kernel and length scale $\ell = 8$, and then train neural networks on each draw.
The results are sets of weights and biases that correspond to smooth functions. 
We then use these weights and biases to define the prior mean $m$ and prior covariance $B$.
We note that draws from this prior result in weights and biases for neural networks whose outputs do not necessarily resemble draws from the Gaussian process we used in the training process.  
\begin{figure}[tb]
    \centering
    \includegraphics[width=0.95\linewidth]{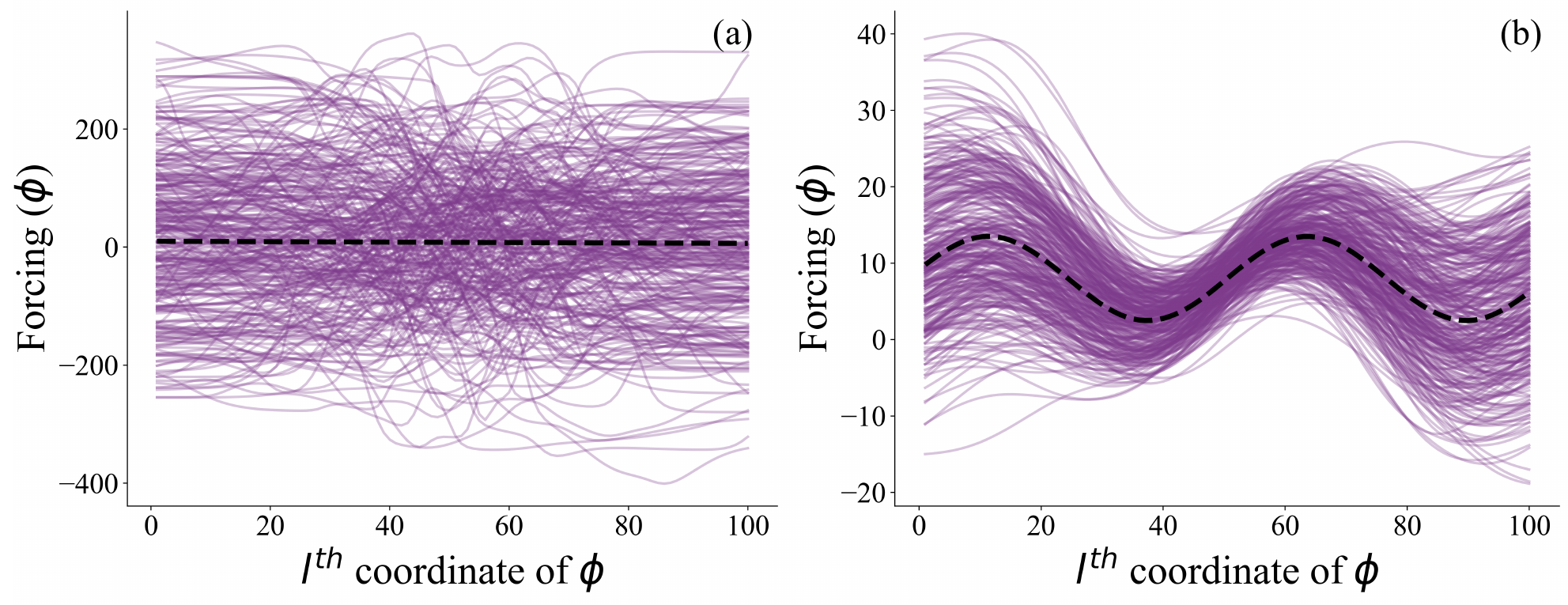}
    \caption{Random samples taken from the prior distributions constructed for the L'96 model with a neural network parameterization.  The purple lines are the functions generated by neural networks whose weights and biases are draws from a Gaussian prior. The black dashed lines are the prior means.
    (a) Samples from the ``uninformative'' prior.  (b) Samples from the ``informative'' prior.}
    \label{fig:nn-priors}
\end{figure}
Figure~\ref{fig:nn-priors}(a) illustrates the outputs (forcings) of neural networks whose weights and biases were drawn from the uninformative prior.
The large spread in the functions illustrates that this prior is uninformative, both about the weights of the neural networks and about the corresponding forcing functions.

As a second contrastive approach, we define a Gaussian directly as a local perturbation of weights from one NN fit to a function that is closer to the true function. 
We define the mean $m$ as the weights and biases of a neural network that we fitted to a randomly translated and compressed version of the true forcing function.
The covariance matrix is a scaled identity matrix $B=0.1I_{n_u}$.
Figure~\ref{fig:nn-priors}(b) shows the outputs of neural networks with weights and biases drawn from this ``informative'' prior. We note that the resulting functions closely resemble the true forcing and, hence this prior is more informative than the prior based simply on the assumption that the forcing be smooth.
We will use the informative and uninformative priors in our numerical examples to illustrate how priors affect the robustness and numerical efficiencies of ensemble methods.
Note that the offline-online approach used here for the informative prior is also discussed by \citeA{PH24} and that similar work on constructing reliable priors for neural networks includes the work of \citeA{KM23} and \citeA{AP23}.

\subsection{Optimal ensemble size of EKI}
\label{sec:opt}

All ensemble methods, except UKI, require that the ensemble size $n_{\text{e}}$ be specified.
This raises the question of how to chose the ensemble size.
In the context of ensemble Kalman filtering, one can rely on works that relate optimal ensemble sizes to unstable and stable directions in chaotic dynamics (see, e.g., \citeA{CB22} and references therein).
For iterative ensemble methods applied to statistical data, however, no such results exist (yet).
Intuitively, a large ensemble size ensures accurate gradient approximations, and the EKI variants can reach the target accuracy with fewer iterations.  
If the ensemble size is small, the gradient approximations are inaccurate and more iterations are required, but each iteration is computationally less costly.
An optimal ensemble size resolves the trade off between fewer iterations with a large ensemble size and more iterations with a smaller ensemble size.
With an  optimal ensemble size defined in this way, we use each EKI variant in a configuration that requires the lowest computational cost to reach a specified target accuracy, where we measure computational cost by the number of forward model runs performed.

To determine the optimal ensemble size we solve the same problem for a range of ensemble sizes and we keep track of how many forward model runs were performed. 
The number of iterations depends on the initial ensemble and the initial conditions for the simulations, which we chose at random, by drawing the initial ensemble from the prior and by perturbing the initial condition.
Thus, the computational cost required is also random. 
We address this issue by performing 100 experiments (at each ensemble size) and then present the average of the number of forward model runs as an indicator of the computational cost.

\begin{figure}[tb]
    \centering
    \includegraphics[width=0.95\linewidth]{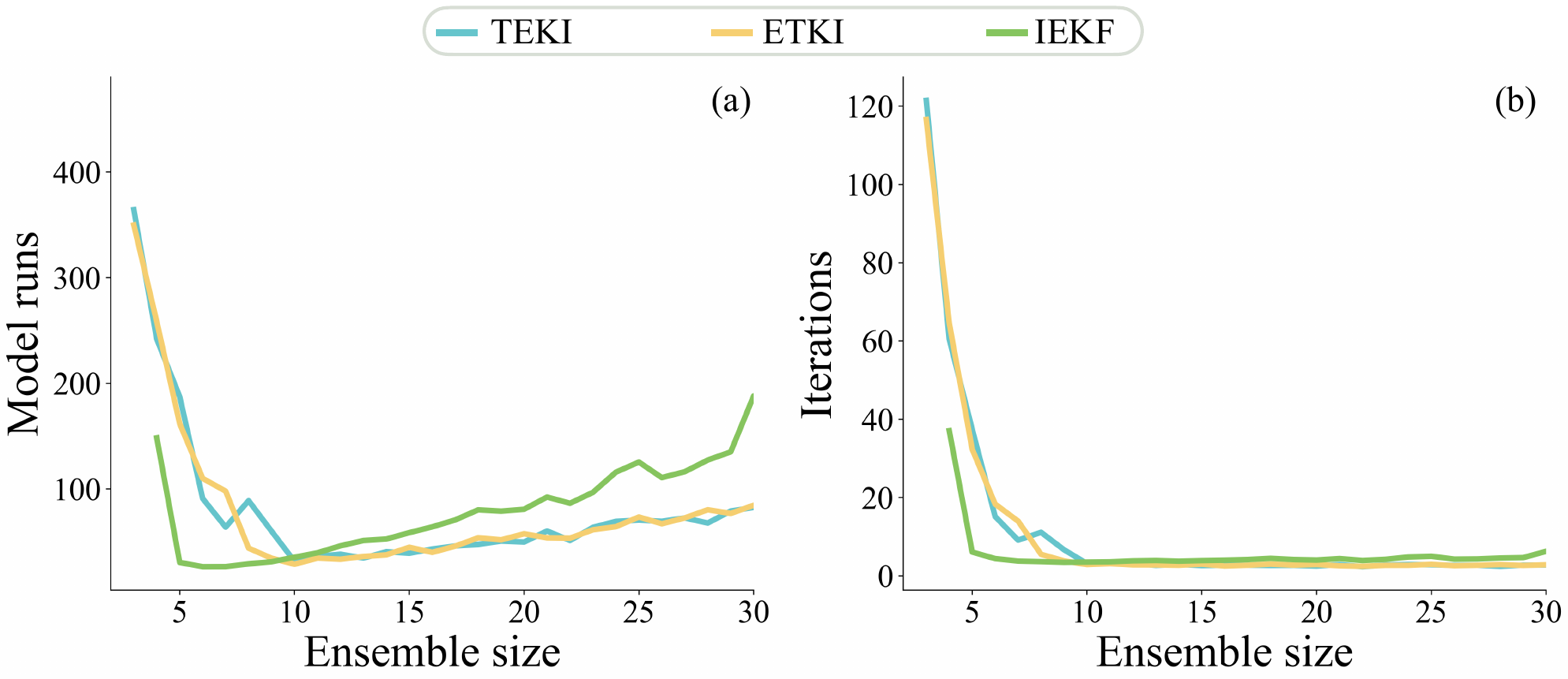}
    \caption{(a) Average number of forward model runs required to reach a target RMSE of one as a function of ensemble size for the L'63 test problem. 
    (b) Number of iterations required to reach a target RMSE of one as a function of ensemble size for the L'63 test problem.
    Shown are results for TEKI (blue), ETKI (yellow), and IEKF (green).
    For UKI, the ensemble size is fixed (Section~\ref{sec:UKI})}
    \label{fig:opt-ens-size}
\end{figure}

An example of the average number of forward model runs required to reach a target accuracy as a function of the ensemble size is shown in Figure~\ref{fig:opt-ens-size}(a).
We present results for the test problem with L'63 dynamics, but the results are qualitatively similar for the other test problems.

While the curves are somewhat noisy, we can clearly see that there is an optimally small ensemble size -- larger ensemble sizes lead to the same result, but at a larger computational cost.
The reason is that the gradient approximations saturate in accuracy once the ensemble size is sufficiently (optimally) large.
Additional ensemble members do not contribute meaningfully to more accurate gradient approximations and inversions with large ensemble sizes still require about the same number of iterations as inversions with the optimal ensemble size.
Figure~\ref{fig:opt-ens-size}(b) shows the average number of iterations required to reach a target accuracy as a function of the ensemble size for the same problem as Figure~\ref{fig:opt-ens-size}(a).
The average number of iterations plateaus as each ensemble method reaches its optimal ensemble size, which indicates that larger ensemble sizes no longer reduce the number of iterations needed to reach the target accuracy.
The additional ensemble members thus simply incur additional computational costs, without speeding up the iteration.

The elevated computational cost for smaller ensemble sizes (smaller than optimal) are partially caused by ensemble failure, i.e., the ensemble was unable to reach the target accuracy within a maximum number of iterations (between 100-200 depending on the configuration of the test problem). 
For the example in Figure~\ref{fig:opt-ens-size}(a), we note that the results for ETKI and TEKI are quite similar (which we observe in other examples as well), and that IEKF has a rather small optimal ensemble size ($n_{{\text{e}},\text{opt}}=6$ for IEKF, compared to $n_{{\text{e}},\text{opt}}=10$ for TEKI and ETKI).
Finally, we note that the graphs can asymptote at different slopes (consider all three graphs in Figure~\ref{fig:opt-ens-size}(a) for relatively large $n_{\text{e}}$).
The reason is that IEKF requires more iterations than TEKI or ETKI, even if the ensemble size is large (IEKF plateaus at a larger number of iterations than TEKI or ETKI in Figure~\ref{fig:opt-ens-size}(b)).
This underlines our definition of optimal ensemble size:
IEKF requires more iterations (for any ensemble size), but can reach the target accuracy with a small ensemble size;
TEKI and ETKI require fewer iterations once the ensemble size is large enough (i.e., equal or larger than the optimal ensemble size), but the resulting optimal number of ensemble members is larger than for IEKF. 
 
\subsection{EKI stopping criteria}
\label{sec:RMSE}

Derivative-based optimization algorithms typically stop the iteration once convergence is detected, e.g., because the loss function does not change much from one iteration to the next. 
We use a different strategy that emphasizes accuracy: Rather than checking for convergence of the ensemble (or its mean), we check if a desired accuracy is reached (independently of convergence). If a desired accuracy is reached, we stop the iteration.
We chose this accuracy-based exit criteria for two reasons.
\begin{enumerate}
    \item
    Accuracy is perhaps most important for GCM calibration.
    Once the model fits the data adequately, we may no longer want to invest computational resources to ensure that the iteration has converged.
    \item
    The various ensemble methods may converge to solutions with different accuracies, which makes a systematic comparison difficult, because one needs to assess, in hindsight, how the relative computational costs compare to the achieved accuracies.
\end{enumerate}
Accuracy-based exit criteria will ensure that each method attempts to achieve the same target accuracy, 
and we can systematically measure the computational expense required to achieve it.
If an ensemble method is unable to achieve the target accuracy, the associated computational cost is large because the iteration continues until a large number of maximum iterations $n_{\max}$ is reached.

We assess the accuracy by the root mean square error (RMSE) defined as
\begin{linenomath*}
\begin{equation}
    \text{RMSE}_j = \frac{1}{\sqrt{n_d}}\left\|R_d^{-\frac{1}{2}}\left(d - \mathcal{F}\left(\bar{u}_j\right)\right)\right\|,
\end{equation}
\end{linenomath*}
where $\bar{u}_j$ is the ensemble mean at iteration $j$.
The EKI iterations stop once the RMSE is below a target RMSE.
A good target is RMSE $=1$, because then the errors in the simulated data are comparable to the errors we expect, as defined by the observation error covariance $R_d$.
In our numerical experiments, we use an RMSE $=1$ as our default choice, but also consider slightly larger target values of RMSE $=1.1$ or $\text{RMSE}=1.2$ to see if we can achieve computational speed up by sacrificing some accuracy.

\subsection{Summary of the experimental setup}
We perform the same systematic numerical experiments for the dynamics\slash parameterizations and resulting forward models described in Section~\ref{sec:time}.
The experimental setup consists of the following steps.
\begin{enumerate}
    \item 
    Estimate the observation error covariance matrix $R_d$ (Section~\ref{sec:DataAndErrorModel}).
    \item 
    Compute the optimal ensemble size for each ensemble method (Section~\ref{sec:opt}).
    \item 
    For each ensemble method, determine how many forward model runs are needed to reach the target RMSE (at optimal ensemble size), using the stopping criteria described in Section~\ref{sec:RMSE}.
\end{enumerate}
The various ensemble methods are all initialized with ensemble members drawn from the prior (Section~\ref{sec:prior}).
We repeat these experiments for different target accuracies to check if we can reduce the computational cost by sacrificing some accuracy.

\section{Numerical experiments}
\label{sec:results}
We report the results of the numerical experiments and focus on the ensemble methods. 
Specifically, we discuss the computational cost, as measured by the number of forward model runs, 
the number of iterations required and the optimal ensemble size.
Variations in these quantities are due to the randomization of the initial ensemble (which affects TEKI, ETKI and IEKF) 
and the random initial condition (which affects all four methods, TEKI, ETKI, IEKF and UKI).
The optimal ensemble size is obtained by averaging over 100 experiments with random initial conditions and initial ensemble draws and, hence, is not random in our setup for TEKI, ETKI and IEKF.
The ensemble size of UKI is always set by the number of unknown parameters (see equation~\eqref{eq:UKIEnsembleSize}).
For all experiments, we consider three different accuracies, corresponding to target RMSEs of 1, 1.1 and 1.2 (higher target RMSE implying lower accuracy).
Optimization results with a derivative-based Levenberg-Marquardt (LM) algorithm are briefly summarized in
Section~\ref{sec:GradientBasedOptimizationResults}.

Our explanations of the results we obtain are detailed, in part because the computational costs are influenced by intertwined effects (optimal ensemble size, number of iterations, informative vs. uninformative priors, number of unknowns, target accurcay). 
On a first read, one may glance over some of the details and refer to the summary in Section~\ref{sec:SummaryOfResults}.

\subsection{Lorenz'63 dynamics} 
\label{sec:L63}
\begin{figure}[tb]
    \centering
    \includegraphics[width=0.95\linewidth]{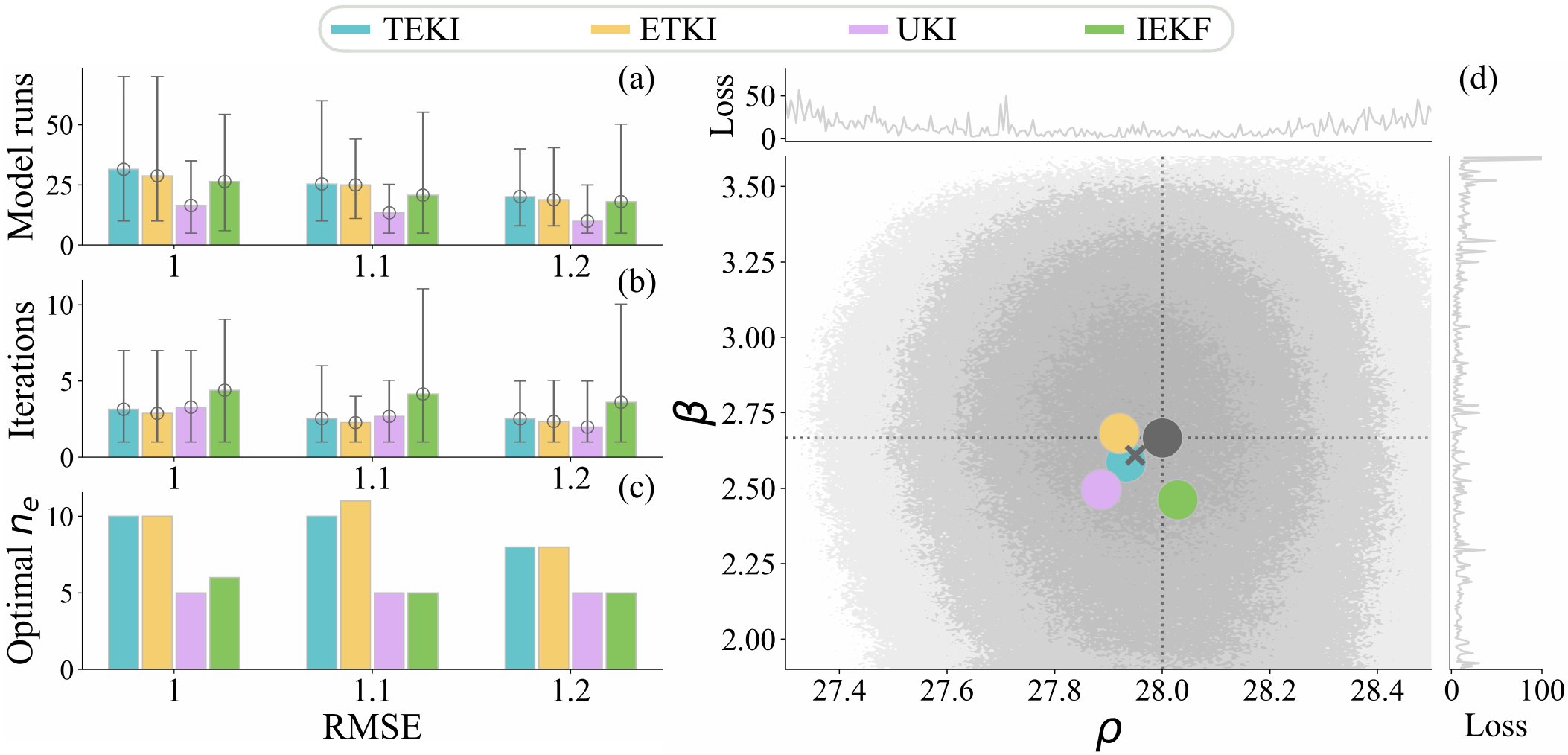}
    \caption{
    Summary of results for estimating two model parameters from statistics of L'63 dynamics.
    (a) Average number of forward model runs required to reach target accuracies of $\text{RMSE}=1$, $\text{RMSE}=1.1$, or $\text{RMSE}=1.2$.
    (b) Average number of iterations required to reach the target accuracies.  
    The error bars in panels (a) and (b) are derived from the 5$^{\text{th}}$ and 95$^{\text{th}}$ percentiles.
    (c) Optimal ensemble sizes for three target accuracies.
    (d) Parameter combinations in the $\rho-\beta$ plane.
    The light gray contours are a 2D histogram of an averaged posterior distribution (darker grays represent regions of higher posterior probability and lighter grays represent regions of lower posterior probability). 
    The ensemble means of TEKI, ETKI, UKI, and IEKF are shown as blue, yellow, pink, and green dots.
    The true parameter pair is a dark gray dot and the expected posterior mode is a dark gray cross. 
    Slices of the loss function obtained by varying $\rho$ or $\beta$ independently and separately are shown at the top and right sides of panel (d).}
    \label{fig:L63}
\end{figure}

We illustrate the results with L'63 dynamics in Figure~\ref{fig:L63}.
Panels (a)-(c) show the number of model runs required to reach the target RMSE, the number of iterations required and the optimal ensemble size -- and we note that the number of model runs is simply the product of the number of iterations and the optimal ensemble size.
For all three target RMSEs, UKI requires the smallest average number of model runs.
The uncertainty (5$^{\text{th}}$ and 95$^{\text{th}}$ percentiles) in the number of required model runs is induced by the randomization over initial conditions, which in turn leads to a varying number of iterations (between two and four).
The ensemble size of UKI is fixed at $n_{\text{e}}=5$, so that the overall cost, which equals the number of iterations multiplied by the ensemble size, amounts to 10-20 forward model runs, depending on the target RMSE and the initial conditions.
We note that the ensemble size of UKI is small ($n_{\text{e}}=2\cdot 2+1=5$) because we only estimate two model parameters ($\rho$ and $\beta$, see~\eqref{eq:UKIEnsembleSize}).

The average cost of IEKF is larger than for UKI, but less than the average cost of TEKI or ETKI.
The low average cost is due to the small optimal ensemble size of IEKF (between $n_{\text{e}}=5$ and $n_{\text{e}}=6$).
IEKF however, is not very robust because the variation in the number of iterations is large, ranging from two to ten for all target RMSEs.
Large variations in the number of iterations then translate into a large variation in the number of model runs.

TEKI and ETKI are comparable in terms of their computational cost, optimal ensemble size, and required number of iterations.
The uncertainties in the number of iterations is also comparable for TEKI and ETKI and it decreases as the target RMSE increases.
We also note that TEKI and ETKI require fewer iterations, but a larger optimal ensemble size than IEKF.
The number of required iterations and the optimal ensemble size, however, can trade off to lead to similar computational costs of these algorithms (on average).

In this problem, UKI is the algorithm of choice because it requires the smallest number of forward model runs for all target RMSEs, even when taking into account uncertainties due to the randomized initial ensembles and initial conditions.
The average cost of IEKF is smaller than the average cost of TEKI or ETKI, but the variation (or uncertainty) in the computational cost of IEKF is larger than the variation of the cost of TEKI or ETKI.
Moreover, the variation in the computational cost of TEKI or ETKI decreases as we increase the target RMSE (lower the specified accuracy).
The large uncertainties associated with the average cost of IEKF make this algorithm less attractive to use than TEKI or ETKI.

We illustrate the parameter estimates of TEKI, ETKI, UKI, and IEKF and the loss landscape of this parameter estimation problem in Figure~\ref{fig:L63}(d).
Specifically, we show the ensemble means of TEKI, ETKI, UKI, and IEKF after the final iteration in the $\rho-\beta$ plane for one of our experiments. 
Since we require that each ensemble method reaches the same target RMSE, the methods find similar model parameters, which are close to the true model parameters.
The gray contours in Figure~\ref{fig:L63}(d) correspond to the expected negative logarithm of the posterior distribution, which we compute by evaluating the loss function (equation~\eqref{eq:loss}) over a grid and averaging.
The true parameter value, as well as the estimates of TEKI, ETKI, UKI or IEKF are all clustered in the region where the averaged loss function is small or, equivalently, where the averaged posterior probability is high.
The top and side panels of Figure~\ref{fig:L63}(d) show slices through the loss function, holding one of two parameters fixed at the true value. 
These slices highlight that the loss function is noisy and characterized by many local minima.
For these reasons, derivative-based optimization converges to the local minimum closest to the starting point of the iteration.
The local minimum identified by derivative-based optimization is often characterized by a large data misfit (see also Section~\ref{sec:GradientBasedOptimizationResults}).

\subsection{Lorenz'96 dynamics: constant forcing} 
\label{sec:L96}
\begin{figure}[tb]
    \centering
    \includegraphics[width=0.95\linewidth]{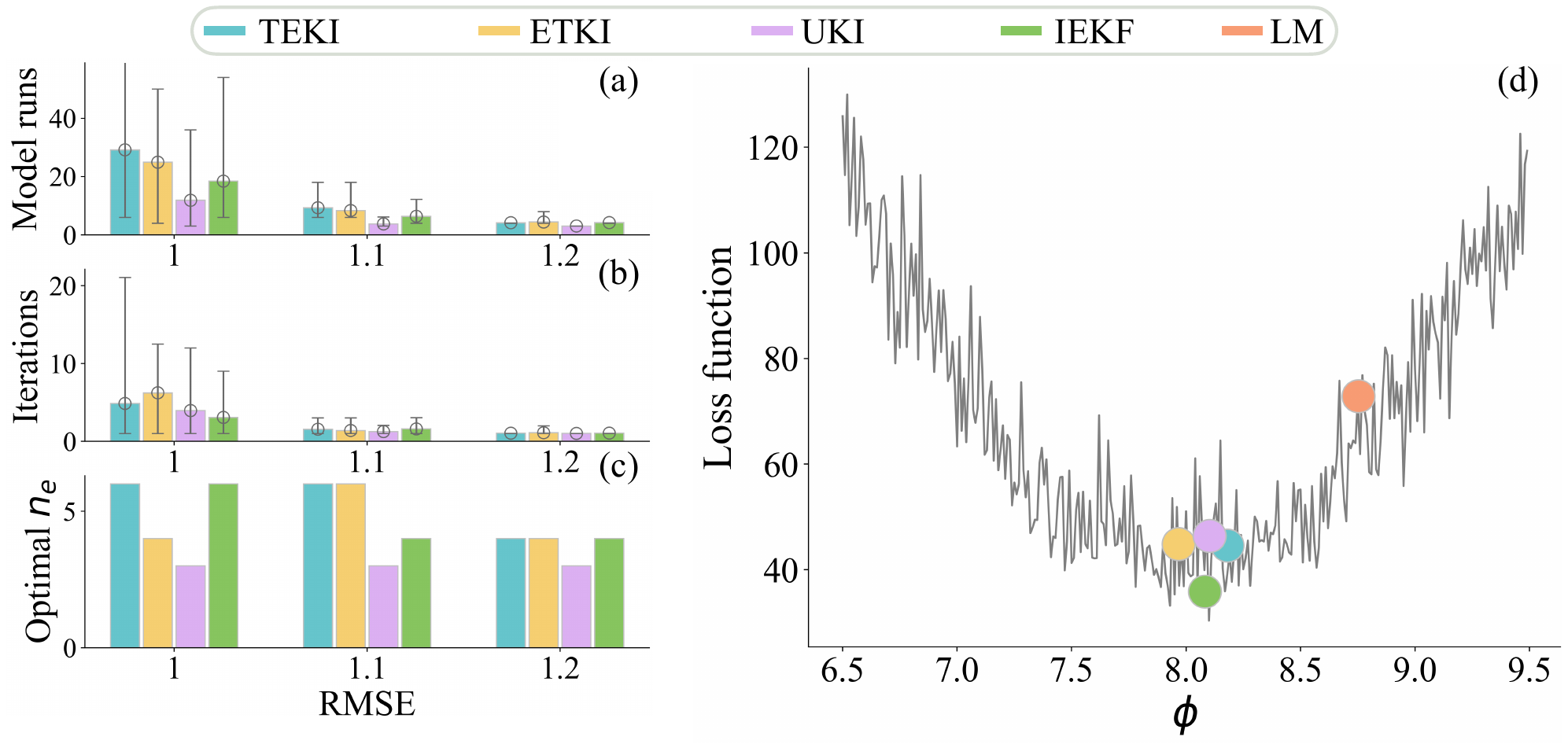}
    \caption{
    Summary of results for the model parameter estimation problem with L'96 dynamics under a constant forcing.
    (a) Average number of forward model runs required to reach target accuracies of $\text{RMSE}=1$, $\text{RMSE}=1.1$, or $\text{RMSE}=1.2$.
    (b) Average number of iterations required to reach the target accuracies.  
    The error bars in panels (a) and (b) are derived from the 5$^{\text{th}}$ and 95$^{\text{th}}$ percentiles.
    Note the truncated $y$-axis in panel (a).
    (c) Optimal ensemble sizes for three target RMSEs.
    (d) Loss function and ensemble means of TEKI, ETKI, UKI, and IEKF for one of our experiments.
    Also shown is the result of a derivative-based optimization via Levenberg-Marquardt, which fails to converge to a useful solution (defined as $\text{RMSE} \leq 1.2$).
    Note that the LM solution does not necessarily fall within a local minima of the depicted loss function because the loss function in itself is random.}
    \label{fig:L96_ConstForcing}
\end{figure}

Results obtained with L'96 dynamics under a constant forcing are summarized in Figure~\ref{fig:L96_ConstForcing}.
All methods struggle to reach a target RMSE of one compared to larger RMSEs, which can be deduced from the large variation in the number of iterations required. In particular, TEKI often requires up to 20 iterations to reach the target RMSE of one. 
All ensemble methods can more reliably reach the target RMSEs 1.1 and 1.2, and the variation in the number of iterations or forward model runs decreases with increasing target RMSE.

Because we only estimate one unknown, the ensemble size of UKI is $n_{\text{e}}=3$, which is smaller than the optimal ensemble sizes we determine for TEKI, ETKI or IEKF (ranging from $n_{\text{e}}=4$ for IEKF and $n_{\text{e}}=4,\dots,6$ for TEKI and ETKI).
The number of iterations required are comparable for all four algorithms and, hence, UKI incurs the lowest computational cost (for all target RMSEs).
For a target RMSE of 1.1, IEKF is more efficient than TEKI or ETKI, taking into account uncertainties due to the initial ensemble draw\slash the initial conditions.
For a target RMSE of 1.2, the efficiency of all four methods is practically independent of the initial ensemble or the initial condition -- we do not observe significant variation in the number of iterations required.
At a lower accuracy (RMSE of 1.2), the computational costs of all four methods are comparable, but UKI incurs the smallest cost due to the small ensemble size (and small number of unknowns).

In summary, we find that all four methods have relatively large variability when the target RMSE is one (desired accuracy is high) and that all methods are more robust and comparable in terms of their computational costs when the target RMSE is larger (RMSE=1.2, i.e., the accuracy is lower). 
At an intermediate accuracy (target RMSE of 1.1), UKI is most efficient and IEKF is more efficient and more reliable than TEKI or ETKI.
Again, we note that the computational efficiency of UKI is tied to the fact that the number of unknowns is so small (we only estimate one parameter).

Figure~\ref{fig:L96_ConstForcing}(d) illustrates the loss landscape (not averaged, target RMSE of one).
The ensemble methods (TEKI, ETKI, UKI, and IEKF) can leverage the large scale convexity of the loss function,
even though locally the loss function is not convex at all (many local minima).
Derivative-based techniques, e.g., Levenberg-Marquardt, cannot utilize the large scale convexity, but rather converge to the local minimum closest to their initialization (see Section~\ref{sec:GradientBasedOptimizationResults}).
In the figure, LM has not converged to a local minimum of the depicted loss, because we only show one instant of the random loss function.
LM converged to a local minimum of another instance of the random loss function.

\subsection{Lorenz'96 dynamics: spatially varying forcing and a grid-based parameterization} 
\label{sec:mL96}
We now consider L'96 dynamics under a spatially varying forcing,
with the forcing function parameterized on the grid, i.e., in the same way the forcing acts on the model that is used to generate the synthetic data.
\begin{figure}[tb]
    \centering
    \includegraphics[width=0.95\linewidth]{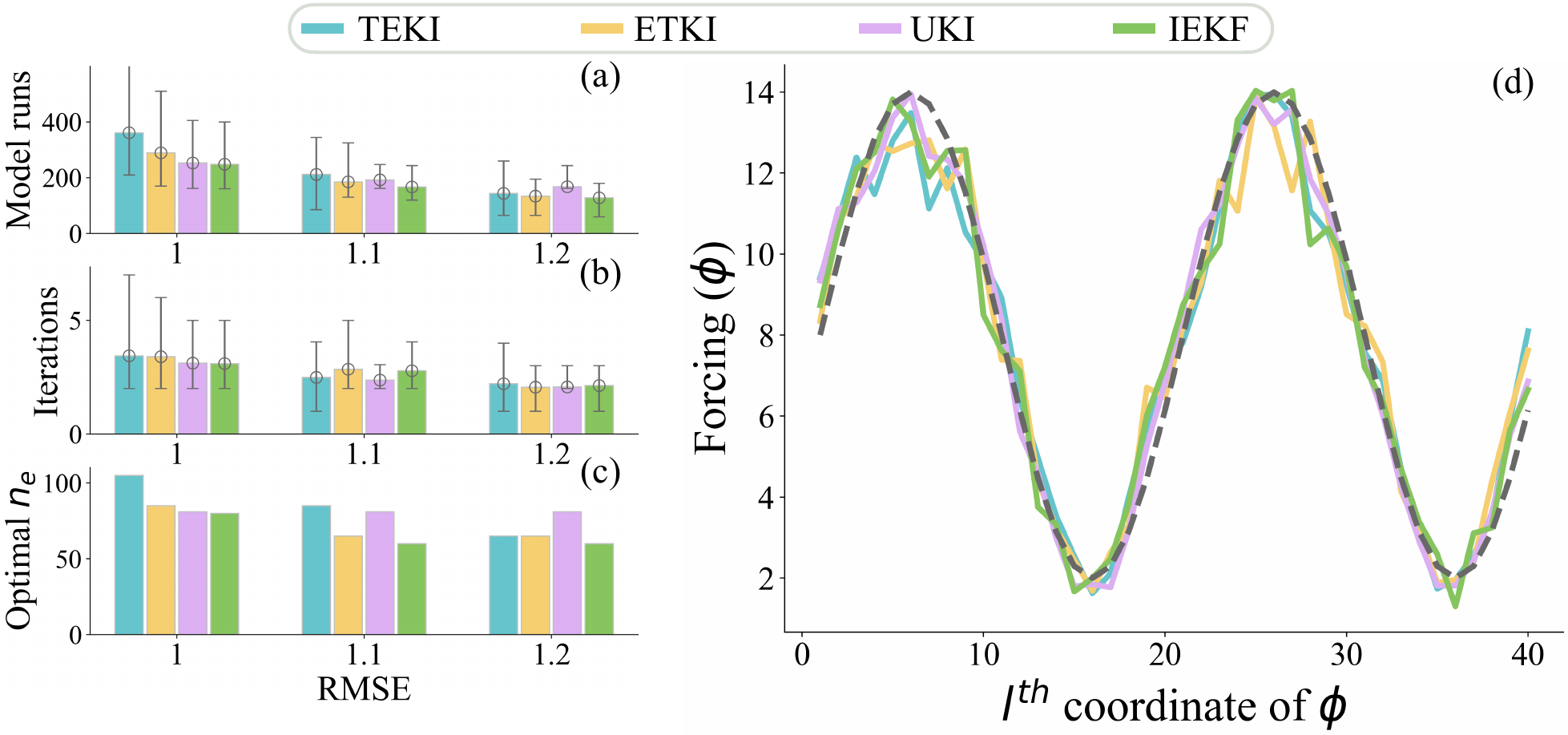}
    \caption{
    Summary of results for L'96 dynamics under a spatially varying forcing, parameterized on a grid.
    (a) Average number of forward model runs required to reach target accuracies of $\text{RMSE}=1$, $\text{RMSE}=1.1$, or $\text{RMSE}=1.2$ (with truncated $y$-axis).
    (b) Average number of iterations required to reach the target accuracies.  
    The error bars in panels (a) and (b) are derived from the 5$^{\text{th}}$ and 95$^{\text{th}}$ percentiles.
    (c) Optimal ensemble sizes for three target RMSEs.
    (d) Spatially varying forcing $\phi_l$ (black, dashed) and estimated grid-based parameterizations obtained by TEKI (blue), ETKI (yellow), UKI (pink), and IEKF (green).}
    \label{fig:L96_VaryForcing_Grid}
\end{figure}
For this grid-based forcing, the number of unknowns is $n_u=40$, so that the ensemble size for UKI is $n_{e}=2\cdot 40+1=81$.
The optimal ensemble sizes of TEKI and ETKI are slightly larger for a target RMSE of 1, but comparable or smaller for target RMSE of 1.1 or 1.2.
The optimal ensemble size of IEKF is smaller than the ensemble size of UKI or TEKI and ETKI for all three target RMSEs.

The number of iterations required on average are comparable for IEKF and UKI, which means that IEKF, in this example, is computationally more efficient than UKI.
For an RMSE target larger than one, IEKF has greater variability in the number of iterations than UKI, but it has smaller ensemble sizes, so the uncertainty in computational cost is comparable between IEKF and UKI. 
TEKI has the largest uncertainty in the number of iterations and overall also the largest computational cost when the target RMSE is set to one.
Similarly, ETKI has larger uncertainty in the number of iterations and a larger computational cost than UKI or IEKF when the target RMSE is set to one.
TEKI and ETKI are comparable at a target RMSE of 1.1, but UKI or IEKF are more efficient.
For a larger target RMSE, TEKI, ETKI and IEKF are comparable in their computational cost and associated uncertainties and all three are more efficient than UKI.

Figure~\ref{fig:L96_VaryForcing_Grid}(d) illustrates the forcing functions the ensemble methods identify from the statistical data.
Since we enforce that all methods reach the same accuracy, all methods generate similar forcing functions,
but we note differences in regions where the forcing is large.
The reason is that a larger forcing implies more chaotic behavior so that the forcing function is less sensitive to the precise value of the forcing, provided that the forcing is sufficiently large.

\subsection{Lorenz'96 dynamics: spatially varying forcing and a neural network parameterization}
\label{sec:mL96nn} 
We now consider L'96 dynamics (dimension $n_z= 100$) under a spatially varying forcing, but for the inversions, we parameterize the forcing function by a neural network (NN) -- mimicking to some extent ML parameterizations in GCMs \cite{VM24}.
The NN parameterization of the forcing has a scalar input (the spatial index $l$) and a scalar output (the forcing $\phi_l$ at index $l$).
We use a shallow NN with 20 neurons and perform online learning to obtain the $n_u=61$ weights and biases of the NN parameterization.
The data are the statistical data of the modified L'96 model (see equation~\eqref{eq:L96_data}).

\subsubsection{Results with an uninformative prior}
\label{sec:mL96nn_uninformative}
With the uninformative prior, UKI can only rarely (about 3\% of the time) reach the target RMSE (one, 1.1 or 1.2), even though UKI has proven to be efficient in the above problems. 
The reason is that the uninformative prior distribution leads to poor initial sigma points and, hence, an initial ensemble with a large data misfit. 
The UKI iteration cannot recover from this poor initial guess and, hence, fails to reach the target RMSE.

\begin{figure}[tb]
    \centering
    \includegraphics[width=0.75\linewidth]{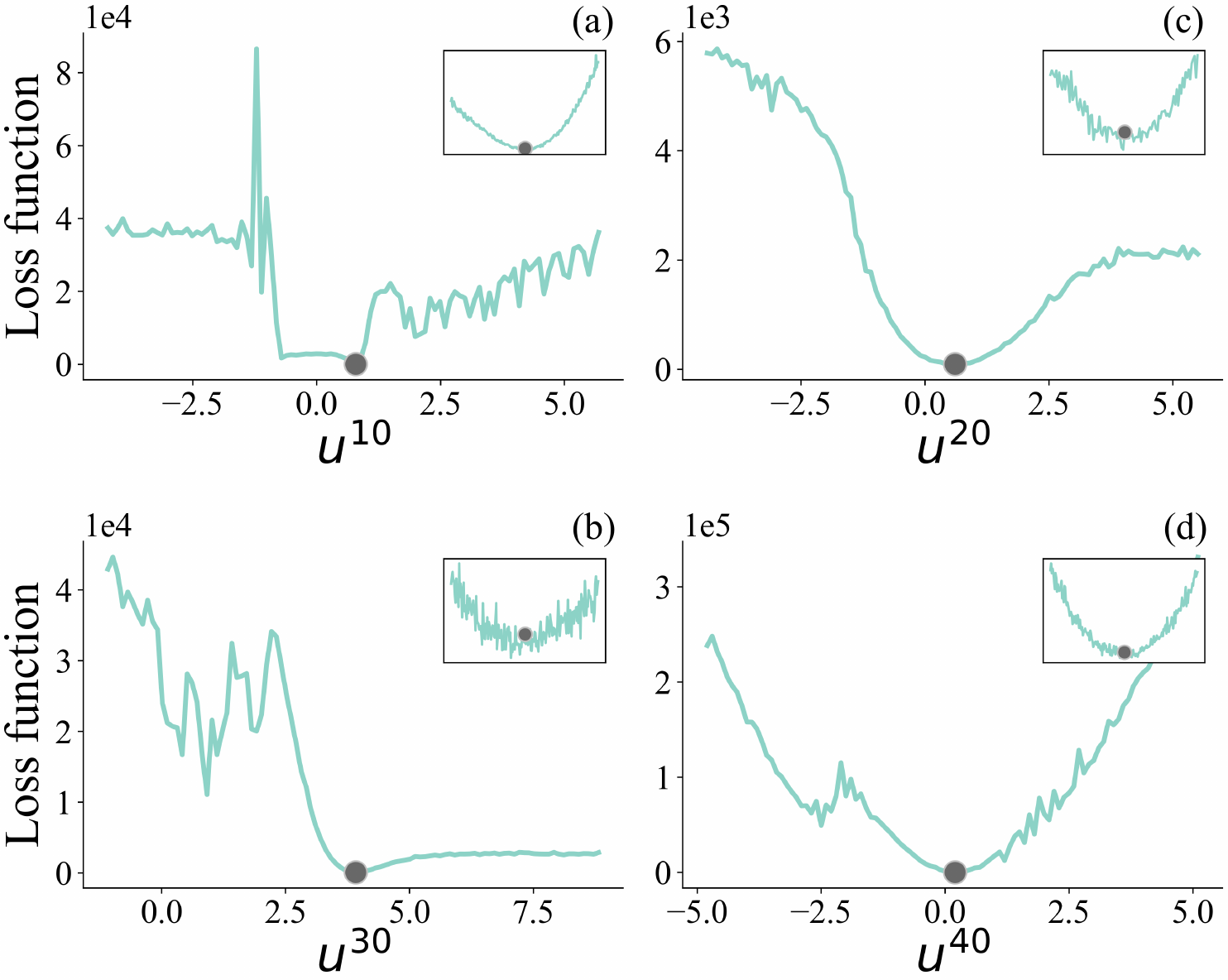}
    \caption{Slices of the loss function defined by the L'96 dynamics with a spatially varying forcing parameterized by a neural network.
    In each slice, all but one of the weights\slash biases is fixed at their true values and one weight is varied.
    The true values of the weights are plotted as dark gray dots.
    A box in each upper-right corner contains a zoomed in portion of the loss function around the true values.
    Near the global minimum, we again observe a large-scale convex structure with noise added due to the statistical data.}
    \label{fig:nnL96-loss}
\end{figure}

The uninformative prior further complicates the loss function landscape, and we do not detect the overall convex loss, corrupted by noise.
Instead, we encounter non-convexity at all scales, as illustrated in Figure~\ref{fig:nnL96-loss}, which shows four slices of the loss function for four neural network weights ($u^{10}, u^{20}, u^{30}, u^{40}$).
Here, the ``true'' values of the networks weights are determined by fitting a neural network offline to the sinusoidal forcing~\eqref{eq:SpatiallyVaryingL96Forcing}. That is, we create a training dataset of input\slash output pairs sampled along the sinusoidal function~\eqref{eq:SpatiallyVaryingL96Forcing} and train the neural network on these data.
We detect that several slices of the loss function are lacking large-scale convex structure.

\begin{figure}[tb]
    \centering
    \includegraphics[width=0.95\linewidth]{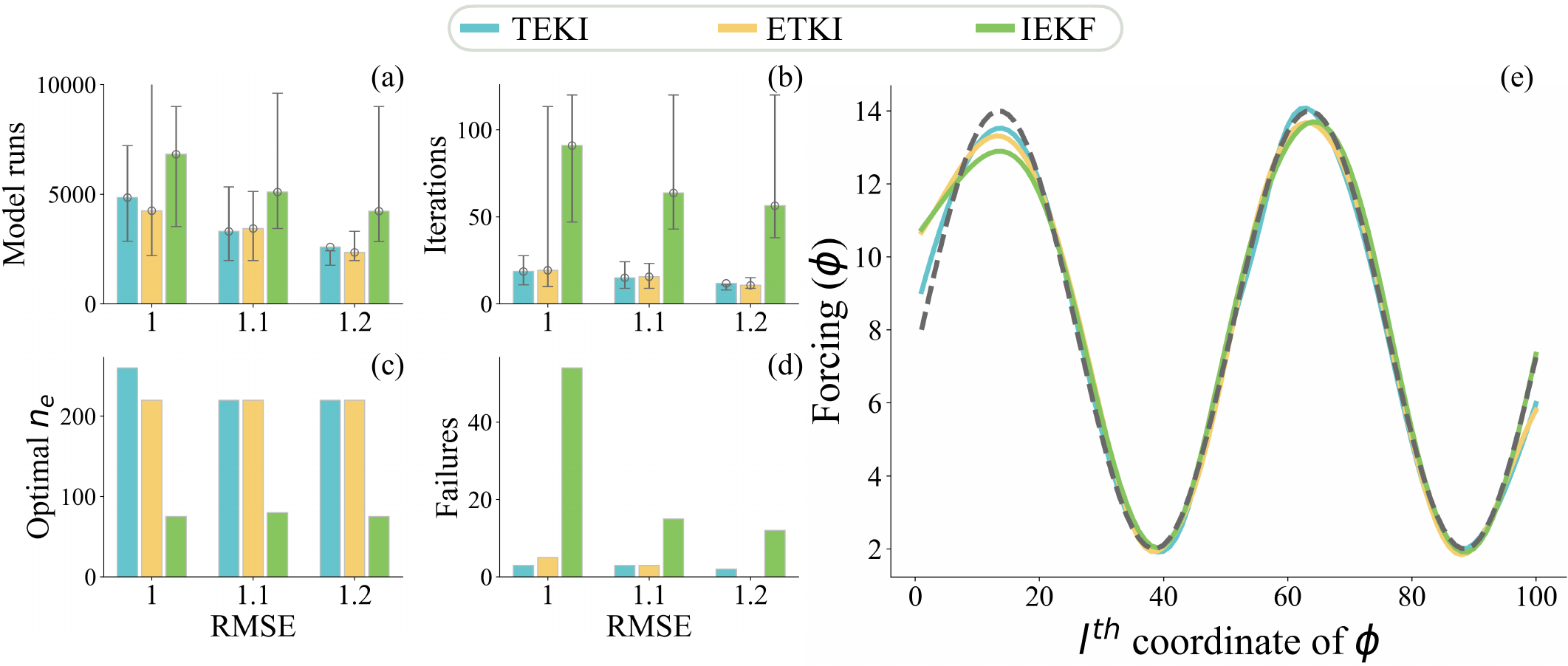}
    \caption{
    Summary of results for L'96 dynamics under a spatially varying forcing, parameterized by a neural network.
    (a) Average number of forward model runs required to reach target accuracies of $\text{RMSE}=1$, $\text{RMSE}=1.1$, or $\text{RMSE}=1.2$ (with truncated $y$-axis).
    (b) Average number of iterations required to reach the target accuracies.  
    The error bars in panels (a) and (b) are derived from the 5$^{\text{th}}$ and 95$^{\text{th}}$ percentiles.
    (c) Optimal ensemble sizes for three target accuracies.
    (d) Total number of ensemble failures (i.e., the RMSE target was not reached within the maximum number of iterations or the algorithm became unstable) over 100 experiments.
    (e) Spatially varying forcing $\phi_i$ (black, dashed) and estimated grid-based parameterizations of TEKI (blue), ETKI (yellow), and IEKF (green). Note: UKI is omitted from these plots because it rarely converged with the uninformative prior.}
    \label{fig:L96_VaryForcing_NN}
\end{figure}

Figure~\ref{fig:L96_VaryForcing_NN} summarizes the results of our numerical experiments with TEKI, ETKI, and IEKF.
First, we note that IEKF has a smaller optimal ensemble size than TEKI or ETKI for all three target RMSEs.
IEKF, however, struggles to converge and often fails to reach the target RMSE within the maximum number of iterations (120).
To improve the stability of IEKF, we used a smaller step size of $\alpha=0.15$ in all experiments with IEKF (larger step sizes lead to even more failures).
For a target RMSE of one, IEKF fails to reach the target RMSE in nearly half of the 100 experiments (Figure~\ref{fig:L96_VaryForcing_NN}(d)).
The number of failures decreases as the target RMSE increases, but the variation in the number of iterations remains high, even for the larger target RMSEs.
In this problem, IEKF is not reliable and computationally more costly than TEKI or ETKI.

TEKI and ETKI are comparable in terms of their optimal ensemble size, number of iterations and number of model runs -- although at a target RMSE of one, ETKI fails to reach the target RMSE more often than TEKI and, therefore, is less reliable.
We further notice that TEKI and ETKI require a relatively large optimal ensemble size (about 200), but only few iterations (about 20).
The reason for the inefficiency of IEKF and UKI is the uninformative prior, which is used very directly by both of these ensemble methods and leads to instability in UKI and a large number of iterations and large variations in the number of iterations for IEKF.
TEKI and ETKI make less explicit use of the prior and, therefore, are more robustly applicable in this problem with an uninformative prior.

Figure~\ref{fig:L96_VaryForcing_NN}(e) illustrates the forcing functions obtained by the ensemble methods and, as before, we see that all methods produce similar results, albeit at very different computational costs (and provided IEKF indeed reaches the target RMSE).
The forcing functions parameterized by the neural networks are smoother than those defined on the grid (Figure~\ref{fig:L96_VaryForcing_Grid}(d)).
The smoother results are expected, because the relatively small neural networks (only 61 weights and biases) with a hyperbolic tangent activation function induce an implicit regularization towards smooth forcing functions.
The fact that TEKI, ETKI and, to some extent, IEKF can find a useful solution by optimizing such a complex loss landscape with a weak prior is encouraging for the possible use of these techniques in the context of GCM calibration.

\subsubsection{Results with an informative prior}
\label{sec:mL96nn_informative}
\begin{figure}[tb]
    \centering
    \includegraphics[width=0.95\linewidth]{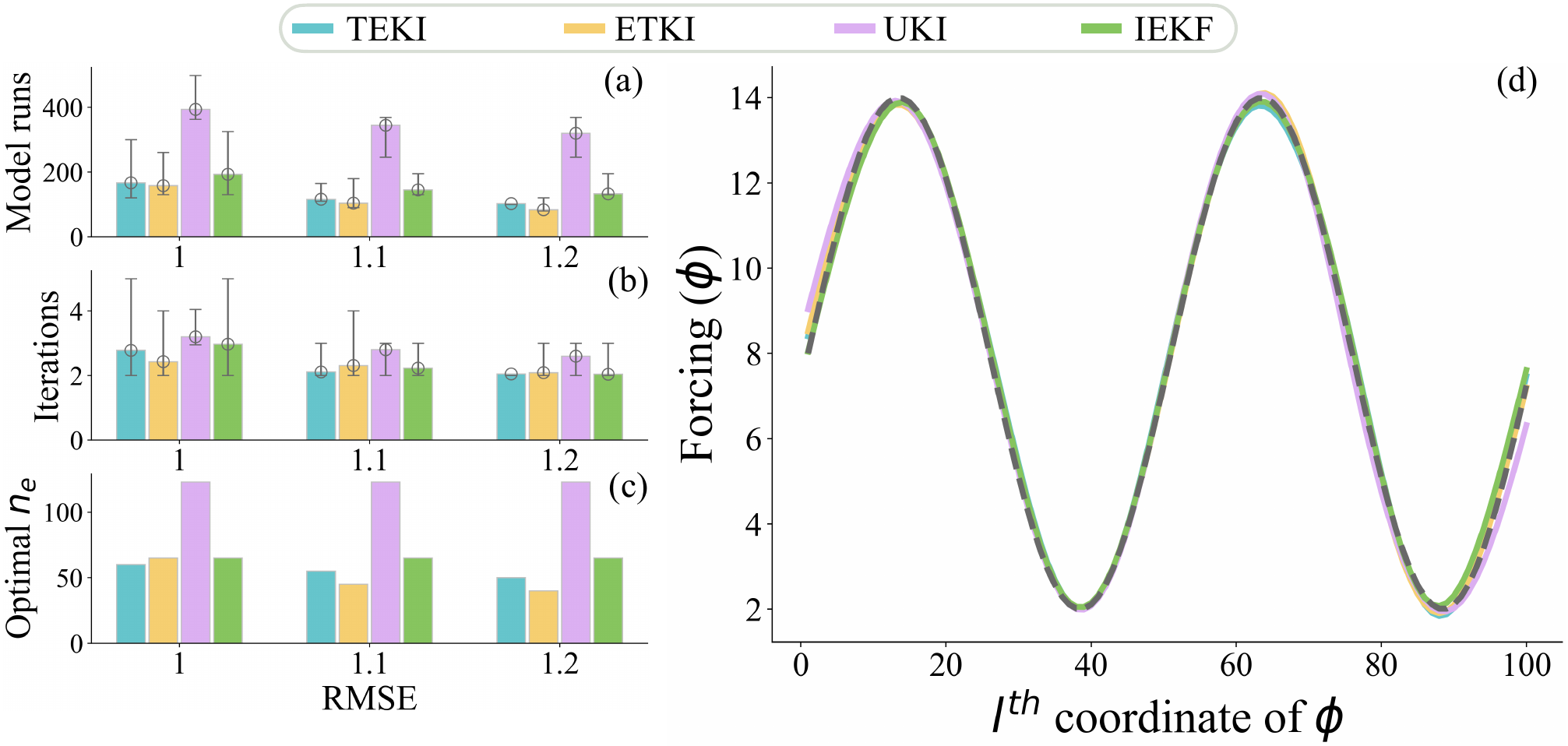}
    \caption{
    Summary of results for L'96 dynamics under a spatially varying forcing, parameterized by a neural network using an informative prior.
    (a) Average number of forward model runs required to reach target accuracies of $\text{RMSE}=1$, $\text{RMSE}=1.1$, or $\text{RMSE}=1.2$.
    (b) Average number of iterations required to reach the target accuracies. 
    The error bars in panels (a) and (b) are derived from the 5$^{\text{th}}$ and 95$^{\text{th}}$ percentiles.
    (c) Optimal ensemble sizes for three target accuracies.
    (d) Spatially varying forcing $\phi_i$ (black, dashed) and estimated grid-based parameterizations of TEKI (blue), ETKI (yellow), UKI (pink), and IEKF (green).}
    \label{fig:L96_VaryForcing_NN_constrained}
\end{figure}

Results we obtain with L'96 dynamics under a spatially varying forcing with a neural network parameterization and an informative prior distribution are summarized in Figure~\ref{fig:L96_VaryForcing_NN_constrained}.
With the informative prior, all four ensemble methods can reach the target RMSEs reliably, which results in relatively small variations in the number of iterations.
Moreover, the informative prior reduces the computational cost of all ensemble methods compared to the experiments with the uninformative prior.
The reason is that the optimal ensemble sizes and the required number of iterations of TEKI, ETKI and IEKF are smaller when the prior is informative 
(compare Figures~\ref{fig:L96_VaryForcing_NN}(c) and~\ref{fig:L96_VaryForcing_NN_constrained}(c)). 
Moreover, the optimal ensemble size, number of iterations, and number of model runs do not vary much from one target RMSE to the next.
Variations in the number of iterations and number of model runs, however, decrease as the target RMSE increases.

For all three target RMSEs, the optimal ensemble sizes of TEKI, ETKI, and IEKF are smaller than the ensemble size of UKI (because the number of unknown parameters is 61, leading to an ensemble size of $n_{\text{e}}=123$ for UKI).
Since UKI requires a larger ensemble size and a larger number of iterations than TEKI, ETKI, or IEKF, UKI is not an efficient algorithm for this problem.
TEKI, ETKI, and IEKF are comparable in their computational cost for all three target RMSEs because their optimal ensemble sizes and number of iterations are comparable.
TEKI, however, has the smallest variation in the computational cost across all three target RMSEs
and IEKF is characterized by a larger number of model runs (on average and when taking uncertainties into account) than TEKI or ETKI.
Thus, even with an informative prior, IEKF can be more computationally expensive than TEKI or ETKI.

Figure~\ref{fig:L96_VaryForcing_NN_constrained}(d) illustrates the forcing functions obtained by the ensemble methods and, as before, we see that all methods produce similar results.
In fact, the forcing estimates for each of the ensemble methods are so similar that the forcing functions have significant overlap. 
Moreover, the forcing functions recovered with the help of an informative prior closely resemble the true forcing function, compared to the grid-based forcing parameterization (Figure~\ref{fig:L96_VaryForcing_Grid}(d)) or the neural network parameterization in the absence of an informative prior (Figure~\ref{fig:L96_VaryForcing_NN}(e)).

\subsection{Summary of results obtained with derivative-based optimization}
\label{sec:GradientBasedOptimizationResults}
Derivative-based optimization via Levenberg-Marquardt (LM) fails to reach the target RMSEs (within the maximum number of iterations) for all numerical experiments we considered and quickly converges to a local minimum near the initialization.
The reason is that the noise in the loss produces many local minima; optimization methods that derive search directions from locally approximated derivatives then get trapped in the local minimum closest to initialization.
To illustrate this effect, we compute the normalized difference of the converged LM solution, $u_*$, to the starting guess $u_0$, defined by
\begin{linenomath*}
\begin{equation}
    \Delta u = \frac{\left\| u_0 - u_*\right\|}{\left\| u_0\right\|},
    \label{eq:normalized difference}
\end{equation}
\end{linenomath*}
where all norms are two-norms.
A small normalized difference indicates that LM converged to a nearby local minimum.
The averaged normalized difference along with the average RMSE of the LM optimization we performed are summarized in Table~\ref{tab:LM_results}.
\begin{table}[t]
    \centering
        \begin{tabular}{cccc}
         Dynamics & Parameterizations (u) & Average RMSE & Average $\Delta u$ \\\hline
         L'63&  Model param. $\rho, \beta$ & 9.856 & $1.316  \times 10^{-1}$\\
         L'96&  Model param. $\phi$ & 3.269 & $1.112\times 10^{-8}$\\
         L'96&  Grid-based $\phi_i$ & 6.529 & $3.467\times10^{-9}$\\
         L'96&  Neural Net (weak prior) & 180.475 & $2.728\times10^{-8}$\\
         L'96&  Neural Net (strong prior) & 17.815 & $1.156 \times10^{-9}$\\
    \end{tabular}
    \caption{Derivative-based optimization via Levenberg-Marquardt (LM) gets trapped in local minima near the starting guess.  
    The result is a small normalized difference $\Delta u$ in~\eqref{eq:normalized difference} and a large RMSE.}
    \label{tab:LM_results}
\end{table}
We conclude that all test problems we consider here are too noisy to be amenable to derivative-based optimization without further modification.

\subsection{Discussion of the results across all five experiments}
\label{sec:SummaryOfResults}
The EKI race with five different test problems does not reveal a clear winner, in the sense that one EKI variant is most efficient in all five test problems.
Our numerical experiments, however, support the following guidelines for the use of EKI variants to provide the most efficient algorithm performance.
\begin{enumerate}
\item \textbf{Use UKI when the problem setting is well-defined and dimensionality is small:}  
\begin{itemize}
    \item UKI wins the race in Sections~\ref{sec:L63} and~\ref{sec:L96} (low-dim, $n_u \leq 10$).
    \item UKI performs worse in Section~\ref{sec:mL96} and loses the race in Section~\ref{sec:mL96nn_informative} (higher dimensions).
    \item UKI failed to converge entirely in Section~\ref{sec:mL96nn_uninformative} (poor prior).
\end{itemize}  
\item \textbf{Use IEKF when the problem setting is well-defined and approximate uncertainty is desired:} 
\begin{itemize}
    \item IEKF did well in Sections~\ref{sec:L63},~\ref{sec:L96},~\ref{sec:mL96}, and~\ref{sec:mL96nn_informative} (good Gaussian priors).
    \item IEKF suffered slow convergence and failures in Section~\ref{sec:mL96nn_uninformative} (poor prior).
    \item Optimal ensemble sizes were always around 1--2$\times n_u$.  
    \item IEKF can provide approximate uncertainty quantification.
\end{itemize}
\item \textbf{Use TEKI/ETKI when the problem setting is less well-defined and robustness is desired:} 
\begin{itemize}
    \item TEKI/ETKI win the race in Sections~\ref{sec:mL96nn_uninformative} and~\ref{sec:mL96nn_informative} (problem most similar to GCM calibration).
    \item TEKI/ETKI performed consistently and robustly across every race.
    \item Optimal ensemble sizes were around 1--3$\times n_u$ in Sections~\ref{sec:L63},~\ref{sec:L96},~\ref{sec:mL96}, and~\ref{sec:mL96nn_informative}, (good Gaussian priors). Optimal ensemble sizes were larger in Section~\ref{sec:mL96nn_uninformative} (poor prior).
    \item ETKI should be used over TEKI for $\dim(y)\gg 10^3$ (linear-algebra formulation). 
\end{itemize}
\end{enumerate}

The fact that TEKI and ETKI performed similarly on all test problems is not a surprise.
Both algorithms solve the same problem in the same way, if we set the ensemble size to infinity.
One may therefore view differences in performance between TEKI and ETKI as being largely driven by uncertainties in the initial ensemble or the initial conditions of the dynamics. 

Our experiments suggest that one can indeed recover model parameters from statistical data using ensemble-based optimization. Indeed, for many test problems, this can be done with all of the methods (although at different computational costs).
The numerical experiments, however, also reveal that IEKF and UKI perform poorly (or fail completely) if the prior is not informative, which may limit their applicability in climate model calibration.

TEKI, ETKI, and IEKF are in principle scalable (by ``scalable,'' we mean that even at a large number of unknowns $n_u$, the number of required forward evaluations remains manageable, e.g., with 500 or less required forward runs) to much higher dimensions than we considered here, in the sense that the ensemble size need not scale with the number of unknowns. 
But, this scaling requires that the covariance estimates within the ensemble iterations are improved via covariance localization or other covariance estimation or sampling error correction methods \cite{TM23,VM24}. 
Without covariance corrections, the minimum number of ensemble members in EnKF (no statistical data) is related to the number of unstable and neutral modes in chaotic dynamics \cite{BC17, GC18}.
No such results are available (yet) for statistical data.
TEKI and IEKF can be localized via traditional, distance-based covariance localization or by applying covariance estimation and sampling error correction techniques to the covariance and cross-covariance matrices.
Localization in transform filters (ETKI) is typically done by considering variables, or sets of variables, independently.
Whether or not that is feasible when the unknowns are weights and biases of neural networks (Section~\ref{sec:mL96nn}) remains to be investigated, but localized transform filters have proven extremely successful in numerical weather prediction.
Thus, whether to use TEKI, IEKF, or ETKI for high-dimensional problems depends on what type of localization strategy is most applicable.
Future work will address questions about scalability of TEKI, ETKI, and IEKF via localization, covariance estimation and sampling error correction.

\section{Summary and conclusions}
\label{sec:conclusions}
Ensemble methods are derivative-free optimization and sampling tools that are successfully applied in several nonlinear Earth science applications, e.g., numerical weather prediction, physical oceanography, and reservoir engineering. 
We consider the use of ensemble methods in the context of climate science problems, specifically global climate model (GCM) calibration, where climate model parameters or parameterizations are estimated from time-averaged data.
The loss functions that arise in this context are noisy and can be non-convex. Ensemble methods can handle such complex loss landscapes because (i) the use of an ensemble instead of a derivative smooths over the noise in the loss function; and (ii) ensemble methods locally linearize statistically, but do not require globally linear models, Gaussian likelihoods or priors, or even, as we have shown, global convexity of the loss function.

Given that there are many variants of ensemble methods in the literature, we have carefully investigated the question: 
``\emph{Which ensemble method variant is most efficient for estimating model parameters from time-averaged data generated by chaotic dynamics}?''
because it is relevant for GCM calibration.
We have designed a set of systematic numerical experiments that allow us to investigate several aspects of the GCM calibration problem: scalability of ensemble methods with the number of unknown parameters, influence of the prior distribution on the numerical efficiency of an ensemble method, and robustness to complex, noisy, non-convex loss landscapes.
The numerical experiments are systematic in that we chose an optimal ensemble size to maximize computational efficiency of each method, define iteration exit criteria to ensure a desired target accuracy, and construct meaningful observation error covariance matrices for time-averaged data.
Such systematic numerical experiments are computationally very expensive to perform with GCMs, even at low resolution.
We designed forward models based on well-known chaotic Lorenz-type dynamics, and considered various configurations of estimating model parameters or model parameterizations.

Being mindful of simplifications implied by the Lorenz-type dynamics, we argued that we can still extract practically useful guidelines from our experiments.
Specifically, our recommendations are that UKI is an efficient option if the number of unknowns is small ($n_u \leq 10$); IEKF is an efficient option if the prior is informative; TEKI and ETKI are the most robust methods, but with a strong prior, IEKF can be computationally more efficient. 
Given the diversity of methods, and consideration of the problem setup, one will likely find a good candidate algorithm.
Moreover, all methods are appropriate in the setting of calibration to statistical data, where derivative-based methods demonstrably fail to produce useful solutions. 
Though not a deep or exhaustive comparison, in the recent age of building auto-differentiable climate model components, this highlights a potential problem that such derivatives may not be usable for parameter calibration to observable statistics.
To improve the existing ensemble methods, research in areas such as removing tight constraints on UKI with high dimensional inputs and implementing ETKI with high dimensional inputs is welcome.

Our numerical experiments do not consider performance enhancements of ensemble methods via covariance localization, inflation, acceleration, and adaptive timestepping though these techniques are critical to the successful application of ensemble methods in high-dimensional problems. 
Our overall framework of performing systematic numerical experiments, however, largely applies to testing such enhancements: We could, for example, re-run the experiments with L'96 dynamics and spatially varying forcing in higher dimensions and with localization and inflation - such experiments are left for follow-up work.
Similarly, it is possible to adjust our framework of numerical experiments to ensemble methods that approximate Bayesian posterior distributions (rather than the optimization methods we focused on).
Many existing methods require additional tuning parameters, and will require new metrics to measure the difference between exact and approximate posterior distributions.

Our relatively simple numerical test problems underline the challenges of hybridizing GCMs with the modern ML techniques, such as introducing data-driven components. 
The loss landscape is always noisy due to the statistical data, but we find neural network parameterizations may induce further complications due to the difficulty of defining reliable (Gaussian) priors that lead to non-convex loss landscapes, and inhibits some ensemble methods from reliable convergence. 
We consider offline-online approaches for the construction of priors for neural network weights, but more work is needed to promote good guidelines.
Our tests suggest that ensemble methods, in particular TEKI and ETKI, can handle such complex loss functions, with and without reliable prior information, and, therefore, remain a promising tool for hybrid GCM calibration.




\section*{Open Research Section}
The software and synthetic data used to generate the results in this paper are uploaded to both Github and Zenodo \cite{rebecca_gjini_2026_20187444}.

\section*{Conflict of Interest Statement}
The authors have no conflicts of interest to disclose.

\acknowledgments
We thank Alberto Carrassi (at the University of Bologna) and an anonymous reviewer for their careful reviews that helped us improve the manuscript.
RG and MM are supported by the US Office of Naval Research (ONR) (grant N00014-21-1-2309). ORAD and TS are supported by Schmidt Sciences, LLC, the U.S. National Science Foundation (grant AGS-1835860), and ONR (grant N00014-23-1-2654).


\bibliography{references}

@article{Wangetal2026,
  author  = {Wang, Libin and Xu, Zhengdao and Zhao, Yan and Tao, Molei},
  title   = {Good regularity creates large learning rate implicit biases: edge of stability, balancing, and catapult},
  journal = {Journal of Machine Learning Research (JMLR)},
  year    = {2026},
  volume  = {26},
  number  = {23-1691},
  url     = {http://jmlr.org/papers/volume26/23-1691/23-1691.pdf}
}

@book{Goodfellow2016,
    title={Deep Learning},
    author={Ian Goodfellow and Yoshua Bengio and Aaron Courville},
    publisher={MIT Press},
    note={\url{http://www.deeplearningbook.org}},
    year={2016}
}

@article{Christopoulos24a,
	author = {Costa Christopoulos and Ignacio Lopez-Gomez and Tom Beucler and Yair Cohen and Charles Kawczynski and Oliver R. A. Dunbar and Tapio Schneider},
	journal = {J. Adv. Model. Earth Sys.},
	pages = {e2024MS004485},
	title = {Online learning of entrainment closures in a hybrid machine learning parameterization},
	volume = {16},
	year = {2024}}

@article{King24a,
	author = {Robert C. King and Laura A. Mansfield  and Aditi Sheshadri},
	journal = {J. Adv. Model. Earth Sys.},
	pages = {e2023MS004163.},
	title = {Bayesian History Matching Applied to the Calibration of a Gravity Wave Parameterization},
	volume = {16},
	year = {2024}}

@article{Deck26a,
	author = {Katherine Deck and Renato K. Braghiere and Alexandre A. Renchon and Julia Sloan and Gabriele Bozzola and Edward Speer and J. Ben Mackay and Teja Reddy and Kevin Phan and Anna L. Gagn{\'e}-Landmann and Yuchen Li and Dennis Yatunin and Andrew Charbonneau and Nat Efrat-Henrici and Eviatar Bach and Shuang Ma and Pierre Gentine and Christian Frankenberg and A. Anthony Bloom and Yujie Wang and Marcos Longo and Tapio Schneider},
	journal = {J. Adv. Model. Earth Sys.},
	pages = {e2025MS005118},
	title = {{ClimaLand}: A Land Surface Model Designed to Enable Data-Driven Parameterizations},
	volume = {18},
	year = {2026}}

@article{CHSV,
author = {Carrillo, J. A. and Hoffmann, F. and Stuart, A. M. and Vaes, U.},
title = {The Mean-Field Ensemble {K}alman Filter: {N}ear-Gaussian Setting},
journal = {SIAM Journal on Numerical Analysis},
volume = {62},
number = {6},
pages = {2549-2587},
year = {2024},
doi = {10.1137/24M1628207}
}

@article{HS22,
  title={Iterated {K}alman methodology for inverse problems},
  author={Huang, Daniel Zhengyu and Schneider, Tapio and Stuart, Andrew M},
  journal={Journal of Computational Physics},
  volume={463},
  pages={111262},
  year={2022},
  publisher={Elsevier}
}

@article{IL13,
doi = {10.1088/0266-5611/29/4/045001},
url = {https://dx.doi.org/10.1088/0266-5611/29/4/045001},
year = {2013},
month = {mar},
publisher = {IOP Publishing},
volume = {29},
number = {4},
pages = {045001},
author = {Marco A Iglesias and Kody J H Law and Andrew M Stuart},
title = {Ensemble {K}alman methods for inverse problems},
journal = {Inverse Problems}
}

@article{CS20,
author = {Chada, Neil K. and Stuart, Andrew M. and Tong, Xin T.},
title = {Tikhonov Regularization within Ensemble {K}alman Inversion},
journal = {SIAM Journal on Numerical Analysis},
volume = {58},
number = {2},
pages = {1263-1294},
year = {2020},
doi = {10.1137/19M1242331},
URL = {https://doi.org/10.1137/19M1242331},
eprint = {https://doi.org/10.1137/19M1242331}
}

@article{CO13,
author = {Chen, Yan and Oliver, Dean S.},
da = {2013/08/01},
doi = {10.1007/s10596-013-9351-5},
id = {Chen2013},
isbn = {1573-1499},
journal = {Computational Geosciences},
number = {4},
pages = {689--703},
title = {Levenberg--{M}arquardt forms of the iterative ensemble smoother for efficient history matching and uncertainty quantification},
ty = {JOUR},
url = {https://doi.org/10.1007/s10596-013-9351-5},
volume = {17},
year = {2013}
}

@article{CC21,
title = {Iterative ensemble {K}alman methods: A unified perspective with some new variants},
journal = {Foundations of Data Science},
volume = {3},
number = {3},
pages = {331-369},
year = {2021},
issn = {},
doi = {10.3934/fods.2021011},
url = {https://www.aimsciences.org/article/id/dbcb5d0f-605d-4752-a353-20f42771f6f3},
author = {Neil K. Chada and Yuming Chen and Daniel Sanz-Alonso}
}

@article{HH22,
doi = {10.1088/1361-6420/ac99fa},
url = {https://dx.doi.org/10.1088/1361-6420/ac99fa},
year = {2022},
month = {oct},
publisher = {IOP Publishing},
volume = {38},
number = {12},
pages = {125006},
author = {Daniel Zhengyu Huang and Jiaoyang Huang and Sebastian Reich and Andrew M Stuart},
title = {Efficient derivative-free {B}ayesian inference for large-scale inverse problems},
journal = {Inverse Problems}
}

@article {BE01,
      author = {Craig H. Bishop and Brian J. Etherton and Sharanya J. Majumdar},
      title = {Adaptive Sampling with the Ensemble Transform {K}alman Filter. {P}art {I}: Theoretical Aspects},
      journal = {Monthly Weather Review},
      year = {2001},
      publisher = {American Meteorological Society},
      address = {Boston MA, USA},
      volume = {129},
      number = {3},
      doi = {10.1175/1520-0493(2001)129<0420:ASWTET>2.0.CO;2},
      pages=      {420 - 436},
      url = {https://journals.ametsoc.org/view/journals/mwre/129/3/1520-0493_2001_129_0420_aswtet_2.0.co_2.xml}
}

@article {TABHW03,
      author = "Michael K. Tippett and Jeffrey L. Anderson and Craig H. Bishop and Thomas M. Hamill and Jeffrey S. Whitaker",
      title = "Ensemble Square Root Filters",
      journal = "Monthly Weather Review",
      year = "2003",
      publisher = "American Meteorological Society",
      address = "Boston MA, USA",
      volume = "131",
      number = "7",
      doi = "10.1175/1520-0493(2003)131<1485:ESRF>2.0.CO;2",
      pages=      "1485 - 1490"
}

@INPROCEEDINGS{WV00,
  author={Wan, E.A. and Van Der Merwe, R.},
  booktitle={Proceedings of the IEEE 2000 Adaptive Systems for Signal Processing, Communications, and Control Symposium (Cat. No.00EX373)}, 
  title={The unscented {K}alman filter for nonlinear estimation}, 
  year={2000},
  volume={},
  number={},
  pages={153-158},
  doi={10.1109/ASSPCC.2000.882463}
  }

@inproceedings{JU97,
author = {Simon J. Julier and Jeffrey K. Uhlmann},
title = {{New extension of the {K}alman filter to nonlinear systems}},
volume = {3068},
booktitle = {Signal Processing, Sensor Fusion, and Target Recognition VI},
editor = {Ivan Kadar},
organization = {International Society for Optics and Photonics},
publisher = {SPIE},
pages = {182 -- 193},
year = {1997},
doi = {10.1117/12.280797},
URL = {https://doi.org/10.1117/12.280797}
}

@book{E09,
author = {Evensen, Geir},
year = {2009},
edition = {Second},
title = {Data assimilation. The ensemble {K}alman filter},
isbn = {978-3-642-03710-8},
publisher = {Springer},
doi = {10.1007/978-3-642-03711-5}
}

@article{E94,
  title={Sequential data assimilation with a nonlinear quasi‐geostrophic model using 
         {M}onte {C}arlo methods to forecast error statistics},
  author={Geir Evensen},
  journal={Journal of Geophysical Research},
  year={1994},
  volume={99},
  pages={10143-10162},
  url={https://api.semanticscholar.org/CorpusID:16213443}
}

@article{E18,
author = {Evensen, Geir},
da = {2018/06/01},
doi = {10.1007/s10596-018-9731-y},
id = {Evensen2018},
isbn = {1573-1499},
journal = {Computational Geosciences},
number = {3},
pages = {885--908},
title = {Analysis of iterative ensemble smoothers for solving inverse problems},
ty = {JOUR},
url = {https://doi.org/10.1007/s10596-018-9731-y},
volume = {22},
year = {2018}}

@article{CO12,
  title={Ensemble randomized maximum likelihood method as an iterative ensemble smoother},
  author={Chen, Yan and Oliver, Dean S},
  journal={Mathematical Geosciences},
  volume={44},
  pages={1--26},
  year={2012},
  publisher={Springer}
}

@article{GO07,
author = {Gu, Yaqing and Oliver, Dean S.},
title = {An Iterative Ensemble {K}alman Filter for Multiphase Fluid Flow Data Assimilation},
journal = {Society of Petroleum Engineers (SPE) Journal},
volume = {12},
number = {04},
pages = {438-446},
year = {2007},
month = {12},
issn = {1086-055X},
doi = {10.2118/108438-PA},
url = {https://doi.org/10.2118/108438-PA},
eprint = {https://onepetro.org/SJ/article-pdf/12/04/438/2561862/spe-108438-pa.pdf}
}

@article{ER13,
title = {Ensemble smoother with multiple data assimilation},
journal = {Computers \&ß Geosciences},
volume = {55},
pages = {3-15},
year = {2013},
note = {Ensemble {K}alman filter for data assimilation},
issn = {0098-3004},
doi = {https://doi.org/10.1016/j.cageo.2012.03.011},
url = {https://www.sciencedirect.com/science/article/pii/S0098300412000994},
author = {Alexandre A. Emerick and Albert C. Reynolds}
}

@article{ER12,
author = {Emerick, Alexandre A. and Reynolds, Albert C.},
da = {2012/06/01},
doi = {10.1007/s10596-012-9275-5},
id = {Emerick2012},
isbn = {1573-1499},
journal = {Computational Geosciences},
number = {3},
pages = {639--659},
title = {History matching time-lapse seismic data using the ensemble {K}alman filter with multiple data assimilations},
ty = {JOUR},
url = {https://doi.org/10.1007/s10596-012-9275-5},
volume = {16},
year = {2012}}

@article{BS14,
author = {Bocquet, M. and Sakov, P.},
title = {An iterative ensemble {K}alman smoother},
journal = {Quarterly Journal of the Royal Meteorological Society},
volume = {140},
number = {682},
pages = {1521-1535},
doi = {https://doi.org/10.1002/qj.2236},
url = {https://rmets.onlinelibrary.wiley.com/doi/abs/10.1002/qj.2236},
eprint = {https://rmets.onlinelibrary.wiley.com/doi/pdf/10.1002/qj.2236},
year = {2014}
}

@article{MW09,
author = {Mor\'{e}, Jorge J. and Wild, Stefan M.},
title = {Benchmarking Derivative-Free Optimization Algorithms},
journal = {SIAM Journal on Optimization},
volume = {20},
number = {1},
pages = {172-191},
year = {2009},
doi = {10.1137/080724083},
URL = {https://doi.org/10.1137/080724083},
eprint = {https://doi.org/10.1137/080724083}
}

@article{GH20,
author = {Garbuno-Inigo, Alfredo and Hoffmann, Franca and Li, Wuchen and Stuart, Andrew M.},
title = {Interacting Langevin Diffusions: Gradient Structure and Ensemble {K}alman Sampler},
journal = {SIAM Journal on Applied Dynamical Systems},
volume = {19},
number = {1},
pages = {412-441},
year = {2020},
doi = {10.1137/19M1251655}
}

@article{GN20,
  title={Affine invariant interacting Langevin dynamics for Bayesian inference},
  author={Garbuno-Inigo, Alfredo and N\"{u}sken, Nikolas and Reich, Sebastian},
  journal={SIAM Journal on Applied Dynamical Systems},
  volume={19},
  number={3},
  pages={1633--1658},
  year={2020},
  publisher={SIAM}
}

@article{SS16,
author = {Schillings, Claudia and Stuart, Andrew},
year = {2017},
month = {02},
pages = {1264–1290},
title = {Analysis of the Ensemble {K}alman Filter for Inverse Problems},
volume = {55},
number = {3},
journal = {SIAM Journal on Numerical Analysis},
doi = {10.1137/16M105959X}
}

@article{SS18, title={Convergence analysis of ensemble {K}alman inversion: the linear, noisy case}, volume={97}, ISSN={0003-6811}, DOI={10.1080/00036811.2017.1386784}, note={Funding by Engineering and Physical Sciences Research Council (EPSRC)}, number={1}, journal={Applicable Analysis: An International Journal}, publisher={Taylor and Francis}, author={Schillings, C. and Stuart, A. M.}, year={2018}, pages={107–123} 
}

@article{CT19,
  title={Convergence Acceleration of Ensemble {K}alman Inversion in Nonlinear Settings},
  author={Neil K. Chada and Xin T. Tong},
  journal={Mathematics of Computation},
  year={2019},
  volume={91},
  number = {335},
  pages={1247-1280},
  url={https://api.semanticscholar.org/CorpusID:207880575}
}

@article{CG21,
title = {Calibrate, emulate, sample},
journal = {Journal of Computational Physics},
volume = {424},
pages = {109716},
year = {2021},
issn = {0021-9991},
doi = {https://doi.org/10.1016/j.jcp.2020.109716},
url = {https://www.sciencedirect.com/science/article/pii/S0021999120304903},
author = {Emmet Cleary and Alfredo Garbuno-Inigo and Shiwei Lan and Tapio Schneider and Andrew M. Stuart}
}

@article{BD22,
author = {Bieli, Melanie and Dunbar, Oliver R. A. and de Jong, Emily K. and Jaruga, Anna and Schneider, Tapio and Bischoff, Tobias},
title = {An Efficient {B}ayesian Approach to Learning Droplet Collision Kernels: Proof of Concept Using “Cloudy,” a New n-Moment Bulk Microphysics Scheme},
journal = {Journal of Advances in Modeling Earth Systems},
volume = {14},
number = {8},
year = {2022},
pages = {e2022MS002994},
doi = {https://doi.org/10.1029/2022MS002994},
url = {https://agupubs.onlinelibrary.wiley.com/doi/abs/10.1029/2022MS002994},
eprint = {https://agupubs.onlinelibrary.wiley.com/doi/pdf/10.1029/2022MS002994},
}

@article{SS20,
author = {Schneider, Tapio and Stuart, Andrew and Wu, Jinlong},
year = {2020},
month = {04},
pages = {},
title = {Learning Stochastic Closures Using Ensemble {K}alman Inversion},
volume = {5},
number = {1},
journal = {Transactions of Mathematics and Its Applications},
doi = {10.1093/imatrm/tnab003}
}

@article{LC22,
author = {Lopez-Gomez, Ignacio and Christopoulos, Costa and Langeland Ervik, Haakon Ludvig and Dunbar, Oliver R. A. and Cohen, Yair and Schneider, Tapio},
title = {Training Physics-Based Machine-Learning Parameterizations With Gradient-Free Ensemble Kalman Methods},
journal = {Journal of Advances in Modeling Earth Systems},
volume = {14},
number = {8},
pages = {e2022MS003105},
doi = {https://doi.org/10.1029/2022MS003105},
url = {https://agupubs.onlinelibrary.wiley.com/doi/abs/10.1029/2022MS003105},
eprint = {https://agupubs.onlinelibrary.wiley.com/doi/pdf/10.1029/2022MS003105},
year = {2022}
}

@article{DLG22, doi = {10.21105/joss.04869}, url = {https://doi.org/10.21105/joss.04869}, year = {2022}, publisher = {The Open Journal}, volume = {7}, number = {80}, pages = {4869}, author = {Oliver R. A. Dunbar and Ignacio Lopez-Gomez and Alfredo Garbuno-Iñigo and Daniel Zhengyu Huang and Eviatar Bach and Jin-long Wu}, title = {Ensemble{K}almanProcesses.jl: Derivative-free ensemble-based model calibration}, journal = {Journal of Open Source Software} }

@Article{SL24,
AUTHOR = {Schneider, T. and Leung, L. R. and Wills, R. C. J.},
TITLE = {Opinion: Optimizing climate models with process-knowledge, resolution, and {AI}},
JOURNAL = {EGUsphere},
VOLUME = {2024},
YEAR = {2024},
PAGES = {1--26},
URL = {https://egusphere.copernicus.org/preprints/2024/egusphere-2024-20/},
DOI = {10.5194/egusphere-2024-20}
}

@article{SB23,
author = {Schneider, Tapio and Behera, Swadhin and Boccaletti, Giulio and Deser, Clara and Emanuel, Kerry and Ferrari, Raffaele and Leung, L. and Lin, Ning and Müller, Thomas and Navarra, Antonio and Ndiaye, Ousmane and Stuart, Andrew and Tribbia, Joseph and Yamagata, Toshio},
year = {2023},
month = {09},
pages = {887-889},
title = {Harnessing {AI} and computing to advance climate modelling and prediction},
volume = {13},
journal = {Nature Climate Change},
doi = {10.1038/s41558-023-01769-3}
}

@article{SL17,
author = {Schneider, Tapio and Lan, Shiwei and Stuart, Andrew and Teixeira, João},
title = {Earth System Modeling 2.0: A Blueprint for Models That Learn From Observations and Targeted High-Resolution Simulations},
journal = {Geophysical Research Letters},
volume = {44},
number = {24},
pages = {12,396-12,417},
year = {2017},
doi = {https://doi.org/10.1002/2017GL076101},
url = {https://agupubs.onlinelibrary.wiley.com/doi/abs/10.1002/2017GL076101},
eprint = {https://agupubs.onlinelibrary.wiley.com/doi/pdf/10.1002/2017GL076101}
}

@article{ST17,
  title={Climate goals and computing the future of clouds},
  author={Schneider, Tapio and Teixeira, Jo{\~a}o and Bretherton, Christopher S and Brient, Florent and Pressel, Kyle G and Sch{\"a}r, Christoph and Siebesma, A Pier},
  journal={Nature Climate Change},
  volume={7},
  number={1},
  pages={3--5},
  year={2017},
  publisher={Nature Publishing Group UK London}
}

@article{DG21,
author = {Dunbar, Oliver R. A. and Garbuno-Inigo, Alfredo and Schneider, Tapio and Stuart, Andrew M.},
title = {Calibration and {U}ncertainty {Q}uantification of {C}onvective {P}arameters in an {I}dealized {GCM}},
journal = {Journal of Advances in Modeling Earth Systems},
volume = {13},
number = {9},
pages = {e2020MS002454},
doi = {https://doi.org/10.1029/2020MS002454},
url = {https://agupubs.onlinelibrary.wiley.com/doi/abs/10.1029/2020MS002454},
eprint = {https://agupubs.onlinelibrary.wiley.com/doi/pdf/10.1029/2020MS002454},
year = {2021},
}

@article{TM23,
doi = {10.1088/1361-6420/accb08},
url = {https://dx.doi.org/10.1088/1361-6420/accb08},
year = {2023},
month = {apr},
publisher = {IOP Publishing},
volume = {39},
number = {6},
pages = {064002},
author = {X T Tong and M Morzfeld},
title = {Localized ensemble {K}alman inversion},
journal = {Inverse Problems}
}

@article {L63,
      author = "Edward N.  Lorenz",
      title = "Deterministic Nonperiodic Flow",
      journal = "Journal of Atmospheric Sciences",
      year = "1963",
      publisher = "American Meteorological Society",
      address = "Boston MA, USA",
      volume = "20",
      number = "2",
      doi = "10.1175/1520-0469(1963)020<0130:DNF>2.0.CO;2",
      pages=      "130 - 141",
      url = "https://journals.ametsoc.org/view/journals/atsc/20/2/1520-0469_1963_020_0130_dnf_2_0_co_2.xml"
}

@article{L95,
  author = {Edward N. Lorenz},
  title = {Predictability: a problem partly solved},
  year = {1995},
  journal = {Proc. ECMWF Seminar on predictability},
  volume = {1},
  pages = {1-18},
  month = {1995},
  publisher = {ECMWF}
}

@article{VM24,
author = {Vishny, David and Morzfeld, Matthias and Gwirtz, Kyle and Bach, Eviatar and Dunbar, Oliver R. A. and Hodyss, Daniel},
title = {High-Dimensional Covariance Estimation From a Small Number of Samples},
journal = {Journal of Advances in Modeling Earth Systems},
volume = {16},
number = {9},
pages = {e2024MS004417},
doi = {https://doi.org/10.1029/2024MS004417},
url = {https://agupubs.onlinelibrary.wiley.com/doi/abs/10.1029/2024MS004417},
eprint = {https://agupubs.onlinelibrary.wiley.com/doi/pdf/10.1029/2024MS004417},
year = {2024}
}

@article{PH24,
  title={Explainable offline-online training of neural networks for parameterizations: A 1{D} gravity wave-{QBO} testbed in the small-data regime},
  author={Pahlavan, Hamid A and Hassanzadeh, Pedram and Alexander, M Joan},
  journal={Geophysical Research Letters},
  volume={51},
  number={2},
  pages={e2023GL106324},
  year={2024},
  publisher={Wiley Online Library}
}

@article{LW22,
  title={The power of (non-) linear shrinking: A review and guide to covariance matrix estimation},
  author={Ledoit, Olivier and Wolf, Michael},
  journal={Journal of Financial Econometrics},
  volume={20},
  number={1},
  pages={187--218},
  year={2022},
  publisher={Oxford University Press}
}

@article{SJ15,
  title={Two-stage filtering for joint state-parameter estimation},
  author={Santitissadeekorn, Naratip and Jones, Christopher},
  journal={Monthly Weather Review},
  volume={143},
  number={6},
  pages={2028--2042},
  year={2015}
}

@article{L44,
 ISSN = {0033569X, 15524485},
 URL = {http://www.jstor.org/stable/43633451},
 author = {Kenneth Levenberg},
 journal = {Quarterly of Applied Mathematics},
 number = {2},
 pages = {164--168},
 publisher = {Brown University},
 title = {A METHOD FOR THE SOLUTION OF CERTAIN NON-LINEAR PROBLEMS IN LEAST SQUARES},
 urldate = {2025-09-19},
 volume = {2},
 year = {1944}
}

@article{M63,
author = {Marquardt, Donald W.},
title = {An Algorithm for Least-Squares Estimation of Nonlinear Parameters},
journal = {Journal of the Society for Industrial and Applied Mathematics},
volume = {11},
number = {2},
pages = {431-441},
year = {1963},
doi = {10.1137/0111030},
URL = {https://doi.org/10.1137/0111030},
eprint = {https://doi.org/10.1137/0111030}
}

@book{NW06,
author = {J. Nocedal and S.J. Wright},
year = {2006},
edition = {Second},
title = {Numerical Optimization},
publisher = {Springer}
}

@inproceedings{KM23,
  title={Imposing functional priors on {B}ayesian neural networks},
  author={Kozyrskiy, Bogdan and Milios, Dimitrios and Filippone, Maurizio},
  booktitle={ICPRAM 2023, 12th International Conference on Pattern Recognition Applications and Methods},
  year={2023}
}

@article{AP23,
author = {Julyan Arbel and Konstantinos Pitas and Mariia Vladimirova and Vincent Fortuin},
title = {{A Primer on Bayesian Neural Networks: Review and Debates}},
volume = {41},
journal = {Statistical Science},
number = {2},
publisher = {Institute of Mathematical Statistics},
pages = {316 -- 353},
year = {2026},
doi = {10.1214/24-STS969},
URL = {https://doi.org/10.1214/24-STS969}
}

@article{AS23,
    author = {Al-Ghattas, Omar and Sanz-Alonso, Daniel},
    title = {Non-asymptotic analysis of ensemble {K}alman updates: effective dimension and localization},
    journal = {Information and Inference: A Journal of the IMA},
    volume = {13},
    number = {1},
    pages = {iaad043},
    year = {2023},
    month = {12},
    issn = {2049-8772},
    doi = {10.1093/imaiai/iaad043},
    url = {https://doi.org/10.1093/imaiai/iaad043},
    eprint = {https://academic.oup.com/imaiai/article-pdf/13/1/iaad043/57003522/iaad043.pdf},
}

@software{rebecca_gjini_2026_20187444,
  author       = {Rebecca Gjini},
  title        = {rgjini/the-eki-race: v0.1.1 [{S}oftware]},
  month        = may,
  year         = 2026,
  publisher    = {Zenodo},
  version      = {v0.1.1},
  doi          = {10.5281/zenodo.20187444},
  url          = {https://doi.org/10.5281/zenodo.20187444},
}

@article{RP13,
  title={Estimating Model Parameters with Ensemble-Based Data Assimilation: A Review},
  author={Juan Jose Ruiz and Manuel Pulido and Takemasa Miyoshi},
  journal={Journal of the Meteorological Society of Japan. Ser. II},
  volume={91},
  number={2},
  pages={79-99},
  year={2013},
  doi={10.2151/jmsj.2013-201}
}

@article{VC25,
   title={Uncertainty quantification for deep learning},
   volume={4},
   ISSN={2634-4602},
   url={http://dx.doi.org/10.1017/eds.2025.10027},
   DOI={10.1017/eds.2025.10027},
   journal={Environmental Data Science},
   publisher={Cambridge University Press (CUP)},
   author={van Leeuwen, Peter Jan and Chiu, Jui-Yuan Christine and Yang, Chen-Kuang Kevin},
   year={2025} }

@Inbook{CB22,
author="Carrassi, Alberto
and Bocquet, Marc
and Demaeyer, Jonathan
and Grudzien, Colin
and Raanes, Patrick
and Vannitsem, St{\'e}phane",
editor="Park, Seon Ki
and Xu, Liang",
title="Data Assimilation for Chaotic Dynamics",
bookTitle="Data Assimilation for Atmospheric, Oceanic and Hydrologic Applications (Vol. IV)",
year="2022",
publisher="Springer International Publishing",
address="Cham",
pages="1--42",
isbn="978-3-030-77722-7",
doi="10.1007/978-3-030-77722-7_1",
url="https://doi.org/10.1007/978-3-030-77722-7_1"
}

@article{PT18,
   title={Stochastic parameterization identification using ensemble Kalman filtering combined with maximum likelihood methods},
   volume={70},
   ISSN={1600-0870},
   url={http://dx.doi.org/10.1080/16000870.2018.1442099},
   DOI={10.1080/16000870.2018.1442099},
   number={1},
   journal={Tellus A: Dynamic Meteorology and Oceanography},
   publisher={Stockholm University Press},
   author={Pulido, Manuel and Tandeo, Pierre and Bocquet, Marc and Carrassi, Alberto and Lucini, Magdalena},
   year={2018},
   month=jan, pages={1442099} 
   }

@article {TA20,
      author = "Pierre Tandeo and Pierre Ailliot and Marc Bocquet and Alberto Carrassi and Takemasa Miyoshi and Manuel Pulido and Yicun Zhen",
      title = "A Review of Innovation-Based Methods to Jointly Estimate Model and Observation Error Covariance Matrices in Ensemble Data Assimilation",
      journal = "Monthly Weather Review",
      year = "2020",
      publisher = "American Meteorological Society",
      address = "Boston MA, USA",
      volume = "148",
      number = "10",
      doi = "10.1175/MWR-D-19-0240.1",
      pages=      "3973 - 3994",
      url = "https://journals.ametsoc.org/view/journals/mwre/148/10/mwrD190240.xml"
}

@INPROCEEDINGS{BC17,
       author = {{Bocquet}, Marc and {Carrassi}, Alberto},
        title = "{Four-dimensional ensemble variational data assimilation and the unstable subspace}",
    booktitle = {EGU General Assembly Conference Abstracts},
         year = 2017,
       series = {EGU General Assembly Conference Abstracts},
        month = apr,
        pages = {13327},
       adsurl = {https://ui.adsabs.harvard.edu/abs/2017EGUGA..1913327B}
}

@article{GC18,
author = {Grudzien, Colin and Carrassi, Alberto and Bocquet, Marc},
title = {Asymptotic Forecast Uncertainty and the Unstable Subspace in the Presence of Additive Model Error},
journal = {SIAM/ASA Journal on Uncertainty Quantification},
volume = {6},
number = {4},
pages = {1335-1363},
year = {2018},
doi = {10.1137/17M114073X},
URL = {https://doi.org/10.1137/17M114073X},
eprint = {https://doi.org/10.1137/17M114073X}
}

@article{WG13,
	author = {Williamson, Daniel and Goldstein, Michael and Allison, Lesley and Blaker, Adam and Challenor, Peter and Jackson, Laura and Yamazaki, Kuniko},
	da = {2013/10/01},
	doi = {10.1007/s00382-013-1896-4},
	id = {Williamson2013},
	isbn = {1432-0894},
	journal = {Climate Dynamics},
	number = {7},
	pages = {1703--1729},
	title = {History matching for exploring and reducing climate model parameter space using observations and a large perturbed physics ensemble},
	ty = {JOUR},
	url = {https://doi.org/10.1007/s00382-013-1896-4},
	volume = {41},
	year = {2013}}

@article{CB20,
author = {Kyle Cranmer  and Johann Brehmer  and Gilles Louppe },
title = {The frontier of simulation-based inference},
journal = {Proceedings of the National Academy of Sciences},
volume = {117},
number = {48},
pages = {30055-30062},
year = {2020},
doi = {10.1073/pnas.1912789117},
URL = {https://www.pnas.org/doi/abs/10.1073/pnas.1912789117},
eprint = {https://www.pnas.org/doi/pdf/10.1073/pnas.1912789117},
}

@incollection{SF18,
       publisher = {Chapman and Hall},
       booktitle = {Handbook of Approximate Bayesian computation},
            year = {2018},
           title = {ABC for Climate: Dealing with Expensive Simulators},
           month = {September},
          editor = {Scott A. Sisson and Yanan Fan and Mark A. Beaumont},
            isbn = {9781439881514},
             url = {https://www.crcpress.com/Handbook-of-Approximate-Bayesian-Computation/Sisson-Fan-Beaumont/p/book/9781439881507},
          author = {Holden, Philip and Edwards, Neil and Hensman, James and Wilkinson, Richard}
}

\end{document}